\documentclass[sigconf]{acmart}
\usepackage{subcaption}
\usepackage{enumitem}
\usepackage{balance}

\AtBeginDocument{%
  }

\copyrightyear{2025}
\acmYear{2025}
\setcopyright{cc}
\setcctype{by}
\acmConference[MICRO '25]{58th IEEE/ACM International Symposium on Microarchitecture}{October 18--22, 2025}{Seoul, Republic of Korea}
\acmBooktitle{58th IEEE/ACM International Symposium on Microarchitecture (MICRO '25), October 18--22, 2025, Seoul, Republic of Korea}\acmDOI{10.1145/3725843.3756111}
\acmISBN{979-8-4007-1573-0/2025/10}

\newcommand{\niparagraph}[1]{\vspace{0pt}\noindent\textbf{#1}}

\begin{document}

\title{Characterizing the Efficiency of Distributed Training:\\A Power, Performance, and Thermal Perspective}



\author{Seokjin Go}
\affiliation{%
  \institution{Georgia Institute of Technology}
  \city{Atlanta}
  \state{GA}
  \country{USA}
}
\email{seokjin.go@gatech.edu}

\author{Joongun Park}
\affiliation{%
  \institution{Georgia Institute of Technology}
  \city{Atlanta}
  \state{GA}
  \country{USA}
}
\email{jpark3234@gatech.edu}

\author{Spandan More}
\affiliation{%
  \institution{Georgia Institute of Technology}
  \city{Atlanta}
  \state{GA}
  \country{USA}
}
\email{smore39@gatech.edu}

\author{Hanjiang Wu}
\affiliation{%
  \institution{Georgia Institute of Technology}
  \city{Atlanta}
  \state{GA}
  \country{USA}
}
\email{hwu419@gatech.edu}

\author{Irene Wang}
\affiliation{%
  \institution{Georgia Institute of Technology}
  \city{Atlanta}
  \state{GA}
  \country{USA}
}
\email{irene.wang@gatech.edu}

\author{Aaron Jezghani}
\affiliation{%
  \institution{Georgia Institute of Technology}
  \city{Atlanta}
  \state{GA}
  \country{USA}
}
\email{ajezghani3@gatech.edu}

\author{Tushar Krishna}
\affiliation{%
  \institution{Georgia Institute of Technology}
  \city{Atlanta}
  \state{GA}
  \country{USA}
}
\email{tushar@ece.gatech.edu}

\author{Divya Mahajan}
\affiliation{%
  \institution{Georgia Institute of Technology}
  \city{Atlanta}
  \state{GA}
  \country{USA}
}
\email{divya.mahajan@gatech.edu}

\renewcommand{\shortauthors}{Go et al.}
\renewcommand{\shorttitle}{Characterizing the Efficiency of Distributed Training: A Power, Performance, and Thermal Perspective}

\begin{abstract}
The rapid scaling of Large Language Models (LLMs) has pushed training workloads far beyond the limits of single-node analysis, demanding a deeper understanding of how these models behave across large-scale, multi-GPU systems.
In this paper, we present a comprehensive characterization of LLM training across diverse real-world workloads and hardware platforms, including NVIDIA H100/H200 and AMD MI250 GPUs. We analyze dense and sparse models under various parallelism strategies -- tensor, pipeline, data, and expert -- and evaluate their effects on hardware utilization, power consumption, and thermal behavior.
We further evaluate the effectiveness of optimizations such as activation recomputation and compute-communication overlap.
Our findings show that performance is not determined solely by scaling hardware capacity. Scale-up systems with fewer, higher-memory GPUs can outperform scale-out systems in communication-bound regimes, but only under carefully tuned configurations; in other cases, scale-out deployments achieve superior throughput. We also show that certain parallelism combinations, such as tensor with pipeline, lead to bandwidth underutilization due to inefficient data chunking, while increasing microbatch sizes beyond a certain point induces bursty execution and peak power excursions that worsen thermal throttling. 
These insights reveal how training performance is shaped by complex interactions between hardware, system topology, and model execution. We conclude by offering recommendations for system and hardware design to improve the scalability and reliability of future LLM systems and workloads.
The source code of this project is available at  \textcolor{ACMDarkBlue}{\url{https://github.com/sitar-lab/CharLLM-PPT}}.

\end{abstract}
\maketitle
\section{Introduction}

Large Language Models (LLMs) have driven datacenters to operate at unprecedented levels of computational demand.
Models such as GPT~\cite{gpt-3}, Llama~\cite{llama3}, Mixtral~\cite{mixtral}, Gemini~\cite{gemini25}, and Claude~\cite{claude} require training across thousands of GPUs connected via high-bandwidth interconnect to maintain throughput.
The shift to large-scale distributed training exposes fundamental bottlenecks in compute efficiency, communication overhead, load balancing, and energy consumption that are not well-captured by single-node analysis, necessitating a systematic characterization of these interactions.

A key challenge in large-scale training is selecting an appropriate parallelism strategy. Each approach --  data parallelism (DP), tensor parallelism (TP), pipeline parallelism (PP), and expert parallelism (EP) as used in Mixture-of-Experts (MoE) models -- offers trade-offs in computation, memory efficiency, and communication patterns.
These trade-offs are further influenced by software-level optimizations such as activation recomputation~\cite{activationrecomputation}, which reduces memory usage at the cost of additional computation, and compute-communication overlap, which improves efficiency but increases hardware resource contention~\cite{ccoverlap}.
Moreover, training performance is highly sensitive to pipeline scheduling and microbatch size. For example, interleaved scheduling can improve utilization in PP workloads, but its effectiveness depends on network depth and synchronization barriers \cite{narayanan2021efficient}. 
Similarly, microbatch size significantly impacts load balancing, activation memory, and synchronization overhead, in ways that vary across parallelism strategies.

As the scale of AI training continues to grow, the design of modern infrastructure is becoming increasingly constrained by power consumption, thermal dynamics, and overall system efficiency~\cite{stojkovic2024dynamollm, tapas, googleBlog, ccoverlap-ispass, ccoverlap-arxiv, wham-patent, wham-arxiv}. Dense GPU clusters often face thermal hotspots, power capping, and frequency throttling, reducing throughput and reliability. To address this, cloud providers invest in airflow optimization, dynamic power budgeting, and workload shaping, efforts typically decoupled from algorithmic choices like parallelism or device placement. Analyzing interactions between training techniques and system constraints is complex due to the diverse design space of models, parallelism strategies, hardware, and software.

While most prior work~\cite{deepspeed, pipedream, alpa, phaze, phaze-patent, pip, piper, fluid, megascale, deepseekv3} evaluates parallelism and optimizations~\cite{flashattention, kvcache, blink, taccl, tacos, democratizingai} primarily from a performance perspective, real-world deployments expose overlooked hardware constraints like thermal imbalance, throttling, and bandwidth limits.
These factors, often underestimated in algorithmic or software-centric designs, can substantially degrade throughput and energy efficiency.
In particular, we observe that hardware-induced variability interacts in complex ways with parallelization strategies.

To address this gap, we present the first comprehensive characterization of training-time interactions between runtime distribution choices and hardware behavior across heterogeneous platforms. Our evaluation spans 32 NVIDIA H200 GPUs, 64 NVIDIA H100 GPUs, and 32 AMD MI250 GPUs. The MI250s have a chiplet-based architecture. 
This diverse testbed enables us to analyze the intersection of training algorithms, runtime behavior, and physical constraints.
We focus on three interrelated dimensions:
\textbf{(a) Interplay between hardware behavior and parallelization strategies},  revealing scenarios where configurations like PP with high micro-batch counts either exacerbate or mask hardware inefficiencies.
\textbf{(b) Trade-offs in optimization techniques}, showing how widely-used techniques such as activation recomputation or compute communication overlap can unintentionally amplify hardware bottlenecks.
\textbf{(c) Thermal and power variability}, highlighting how asymmetric GPU throttling disrupts synchronization and skews performance.
This characterization enables us to uncover several key insights. 
\begin{itemize}[leftmargin=*, itemsep=1pt, parsep=0pt, topsep=1pt, partopsep=1pt]
    \item \textbf{Scale-up vs. scale-out performance}: We observe that while scale-out systems (64$\times$H100) generally offer higher performance due to greater aggregate compute, scale-up systems (32$\times$H200) outperform in communication-heavy regimes,  but only under carefully optimized configurations. This sensitivity to the parallelism strategy shows that hardware density alone is insufficient, motivating a co-design approach across model architecture, system topology, and parallelism strategy to fully exploit the potential of emerging high-capacity accelerator platforms.
    
    \item \textbf{Communication inefficiencies in TP+PP}: We observe that the combined use of tensor and pipeline parallelism results in under-utilization of PCIe bandwidth. This is due to sparse SendRecv calls that lack data chunking, leading to inefficient bandwidth usage and higher communication latency. These findings highlight the need for topology-aware collectives that adapt communication patterns to the underlying network layout, ensuring efficient bandwidth utilization and minimizing resource contention.

    \item \textbf{Limits of microbatch scaling:} Increasing the micro-batch size beyond its optimal point, despite having more memory to support this larger size, reduces training efficiency due to communication bandwidth saturation, diminishing compute returns, and more bursty execution patterns. This, in turn, elevates peak power draw and chip temperatures, intensifying thermal throttling and its impact on performance.

    \item \textbf{Thermal imbalance and GPU throttling}: We observe that thermal throttling due to physical placement of nodes and servers introduces significant performance variability, especially when software assumes uniform application of optimizations. Rather than limiting all devices to accommodate the hottest GPUs, a more effective strategy is to more aggressively utilize cooler GPUs, thereby improving throughput without exacerbating overall thermal imbalance.

\end{itemize}

These insights stem from second-order effects like thermal and power imbalance across GPUs, caused by localized cooling inefficiencies or node-level power limits, which lead to execution skew and reduced throughput, especially in synchronization-heavy strategies like tensor and data parallelism.
Even within the same GPU model, hardware characteristics such as thermal behavior and throttling vary across physical environments and can significantly impact performance.
Network latency, frequency scaling, and temperature fluctuations further compound these issues, revealing a tight coupling between system-level dynamics and algorithmic behavior.
\textit{As such, our findings expose a key gap in current ML systems research: parallelism techniques are rarely co-designed with awareness of real-world hardware variability. As LLM training pushes systems to their physical limits, it is crucial to adopt a full-stack perspective that goes beyond algorithmic metrics.}

In one instance during our study, a node-level power failure caused GPUs to run more than 4$\times$ slower, creating severe stragglers that disrupted the entire training pipeline. 
Even without such failures, communication efficiency is highly sensitive to how model parallelism maps onto hardware topology; poor placements amplify latency and bandwidth contention, especially at larger microbatch sizes.
Microbatch size, while useful for smoothing GPU latency variance, also increases communication volume, and its effect varies non-linearly across parallelization strategies.
These results reinforce the need for strategy-aware, topology-conscious tuning of system parameters to ensure efficient and stable LLM training.

Finally, we observe that optimization techniques such as activation recomputation, interleaved pipeline scheduling, and communication compute overlap interact in non-trivial ways with underlying hardware behavior. 
For instance, overlapping communication with computation can mask latency but may also extend kernel execution time due to resource contention. Likewise, activation recomputation eases memory pressure but incurs higher compute overhead, especially in deeply pipelined setups.
Overall, our findings highlight the importance of co-optimizing parallelism strategy and system-level execution--with awareness of both algorithmic structure and hardware behavior--to achieve robust, efficient, and scalable LLM training performance.
As such, this paper makes the following key contributions:
\begin{itemize}[leftmargin=*, itemsep=1pt, parsep=0pt, topsep=1pt, partopsep=1pt]
\item We characterize large-scale LLM training on modern datacenter hardware using fine-grained profiling and system-level metrics, highlighting the interplay between software parallelism and hardware behavior.
\item We analyze tensor, pipeline, data, and expert parallelism across diverse models, quantifying their effects on compute throughput, communication efficiency, and hardware bottlenecks.
\item We study key training optimizations--activation recomputation, interleaved scheduling, and compute-communication overlap--and reveal their interactions with power, thermal behavior, and resource contention.
\item We compare three state-of-the-art GPU platforms, including chiplet-based designs, and uncover non-obvious trade-offs in scaling behavior, thermal limits, and communication patterns that influence system efficiency.
\end{itemize}
\section{Background and Related Work}~\label{sec:background}
\vspace{-3ex}
\subsection{Distributed Machine Learning}
\label{subsec:distributedml}
\begin{figure}
    \centering
    \includegraphics[width=0.9\linewidth]{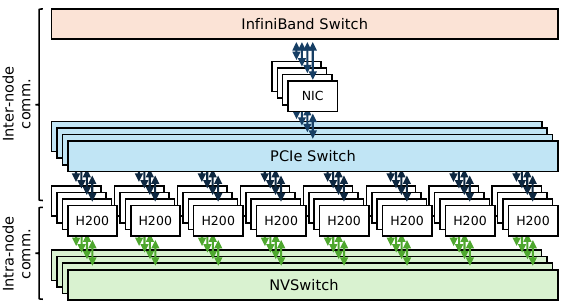}
    \vspace{-2ex}
    \caption{Network topology of four NVIDIA HGX H200 nodes. Intra-node GPUs connect via NVLink/NVSwitch; inter-node communication uses PCIe and InfiniBand.
    }
    \label{fig:network}
    \vspace{-3ex}
\end{figure}

\niparagraph{Infrastructure and runtime systems.} The scale of modern machine learning~\cite{gpt-3, llama3, mixtral, claude} has led to the adoption of large-scale distributed training across multi-node, multi-GPU datacenters~\cite{deepspeed, shoeybi2019megatron}.
These setups, such as the example illustrated in Figure~\ref{fig:network}, rely on domain-specific stacks~\cite{pytorch, tensorflow, tabla, dnnweaver:micro, cosmic:micro}, scalable frameworks such as Megatron~\cite{shoeybi2019megatron}, DeepSpeed~\cite{deepspeed}, and NeMo~\cite{kuchaiev2019nemo}, which introduce techniques for parallelism, memory management, and throughput optimization~\cite{taccl, pipedream, flashattention, deepspeed, rajbhandari2020zero, piper}.
Training often spans thousands of GPUs connected via high-bandwidth interconnects like NVLink, NVSwitch~\cite{nvlink}, and InfiniBand~\cite{infiniband}, stressing infrastructure across compute, communication, power, and thermal limits.

\niparagraph{Parallelism strategies.} Parallelism strategies distribute models across devices with trade-offs in memory, interconnect, and utilization. Data Parallelism (DP)\cite{dataparallel,dataparallel1} replicates the model but incurs gradient sync overhead. FSDP\cite{fsdp} shards parameters to reduce memory and enable larger models. Tensor Parallelism (TP)\cite{shoeybi2019megatron} splits computation across GPUs but demands high bandwidth. Pipeline Parallelism (PP)\cite{pipedream, gpipe, pipedream2bw} stages the model to save memory and overlap compute, but can suffer from stalls. Expert Parallelism (EP)~\cite{deepspeedmoe, megascaleinfer, hybridep} activates subsets of experts, introducing load imbalance and expensive all-to-all communication.

\niparagraph{Device placement.}  Modern LLM training often uses hybrid parallelism, combining DP, TP, PP, and EP tailored to model architecture, memory constraints, and hardware topology.
Prior work on device placement and strategy search optimizes parameters like parallel widths, pipeline depth, and expert parallel configuration~\cite{alpa, flexflow, piper, phaze, topoopt}, considering factors such as kernel latency, network bandwidth, and memory footprint.
However, the effectiveness of these strategies can be limited by system-level constraints including thermal throttling, power capping, and interconnect bottlenecks.
As LLM training pushes infrastructure limits, it becomes essential to extend parallelism strategy design to incorporate system-level considerations, including cooling efficiency, power delivery constraints, and topological layout.
This highlights a critical gap: parallelism and placement strategies are rarely co-designed with awareness of physical infrastructure constraints. In this work, we bridge this gap by empirically characterizing how power, thermal behavior, and interconnect contention impact distributed LLM training, uncovering bottlenecks and challenging assumptions. 
This motivates the need for full-stack approaches that combine parallelism-aware scheduling with system-level modeling.

\niparagraph{Network search and communication collective optimization.}
Previous works leverage underlying network topology information to design communication patterns like all-to-all, ring, or tree-based collectives~\cite{blink, taccl, tacos, sccl}, with some efforts optimizing physical topologies for various models and communication patterns, including all-reduce~\cite{topoopt}.
In contrast, our characterization emphasizes the need to co-optimize communication collectives and network topology alongside parallelism strategies and system-level factors, revealing a broader set of correlations than previously explored.

\subsection{Related Work}\label{subsec:relatedwork}

\niparagraph{Power and thermal aware LLM serving.}
There has been growing attention to power- and thermal- aware execution, especially for LLM inference workloads. Systems like DynamoLLM\cite{stojkovic2024dynamollm} and TAPAS\cite{tapas} propose energy-efficient model placement and resource allocation strategies that leverage power headroom and thermal slack. Splitwise~\cite{splitwise} explores adaptive model partitioning for inference with power variability between the prefill and decode phases.
These works focus on LLM serving, where latency and energy efficiency are primary constraints, with dynamic decisions based on fluctuating query loads and thermal profiles.
In contrast, our work focuses on the training of large-scale LLM systems, where workloads are synchronous, tightly coupled, and far more resource-intensive. Unlike inference, training performance is influenced not only by latency and energy but also by inter-GPU synchronization, communication-compute balance, and overall throughput.

\niparagraph{System and runtime characterization.}
Prior works have explored various aspects of system characterization across hardware and software stacks. Studies on memory hierarchies~\cite{microbenchmarking} and interconnect behavior~\cite{tartan, realGPUNoc} provide insights into GPU microarchitecture, while others~\cite{wang2023understanding, xfaas, characterizationrpc, googlehyperscale, fae, hotline, hotline-patent} focus on cloud-scale systems for distributed services, serverless architectures, and remote procedure calls.
Recent studies have characterized LLM training workloads in datacenters, examining scheduling, hardware failures, and learning dynamics across checkpoints~\cite{charllm}. 
Pythia~\cite{pythia} enables detailed analysis of LLM training dynamics through regular checkpoints and controlled replication across scales.
AxoNN~\cite{democratizingai} and MegaScale~\cite{megascale} characterized performance and scalability of state-of-the-art parallelism and optimization techniques at large-scale training clusters with thousands of GPUs.
While these works offer insights into large-scale training behavior, they primarily focus on software or aggregate metrics. 
In contrast, our work presents the first system-level characterization of distributed LLM training, analyzing performance, power, and thermal implications of parallelism strategies and optimization techniques across hardware platforms to unveil trade-offs affecting real-world efficiency.

\section{Infrastructure Setup}

\subsection{Software Libraries and Workloads}\label{subsec:setup}

\niparagraph{Measurement and telemetry.} Training is conducted using PyTorch 2.6.0~\cite{imambi2021pytorch}. Execution traces are collected with the Chakra Profiler~\cite{sridharan2023chakra}, and energy measurements are captured using a modified version of Zeus~\cite{you2023zeus}. Our Zeus extension enables fine-grained monitoring of component-level metrics, including core and memory temperatures, per-component utilization, and PCIe bandwidth, allowing detailed analysis of thermal behavior, resource usage, and system dynamics. To ensure stable measurements, we discard the first 10 iterations for warm-up, during which GPU temperatures and activity stabilize, then profile subsequent iterations in detail.

\niparagraph{Workloads, distribution strategies, and optimizations.} We evaluate five representative LLMs (see Table~\ref{tab:evaluated-models}), spanning both dense and Mixture-of-Experts (MoE) architectures, with model sizes ranging from 30B to 175B parameters. 
All experiments use the Pile~\cite{pile} dataset, FP16 or BF16 precision, and a global batch size of 128.
For LoRA finetuning and inference tasks, we use the PubMedQA~\cite{pubmedqa} dataset.
We vary tensor, pipeline, data, and expert parallelism configurations based on the model and hardware setup. For each model and cluster, we determine the minimal total model parallelism (Tensor $\times$ Pipeline $\times$ Expert) required to fit within GPU memory, then explore valid configurations, limiting tensor parallelism to within-node execution to reduce communication costs.
We also evaluate Fully-Sharded Data Parallelism (FSDP) in conjunction with tensor parallelism (i.e., TP8+FSDP4), reflecting common 2D parallelism usage in large-scale training~\cite{simplefsdp,fsdp,2dparallellightning,pytorchtp}.
Standalone FSDP was not evaluated because of memory constraints arising from increased microbatch sizes.

\begin{table}[t]
    \centering
    \caption{Evaluated model configurations}
    \vspace{-0.7em}
    \resizebox{0.8\columnwidth}{!}{
    \begin{tabular}{ccc}
        \toprule
        \textbf{Model} & \textbf{Type} & \textbf{Parameter Size} \\
        \midrule
        GPT3-175B~\cite{gpt-3} & Dense & 175B   \\
        GPT3-30B~\cite{gpt-3} & Dense & 30B \\
        Llama3-70B~\cite{llama3} & Dense & 70B \\
        Llama3-30B~\cite{llama3} & Dense & 30B \\
        Mixtral-8x22B~\cite{mixtral} & Mixture-of-Experts & 141B \\
        Mixtral-8x7B~\cite{mixtral} & Mixture-of-Experts & 47B \\
        \bottomrule
    \end{tabular}
    }
    \label{tab:evaluated-models}
\end{table}

\begin{table}[t]
    \centering
    \caption{Evaluated parallelism and optimization techniques. Arrows indicate impact on training time (Perf), memory usage (Memory), and communication intensity (Comm): $\uparrow$/$\downarrow$ for increase/decrease, $\uparrow$$\uparrow$/$\downarrow$$\downarrow$  for stronger effects, and (–) for negligible or mixed impact.}
    \vspace{-0.7em}
    \resizebox{0.85\columnwidth}{!}{
    \begin{tabular}{ccccc}
        \toprule
        \textbf{Technique} & \textbf{Abbr} & \textbf{Perf} & \textbf{Memory} & \textbf{Comm} \\
        \midrule
        Tensor Parallelism & TP & $\downarrow$$\downarrow$ & $\downarrow$ & $\uparrow$$\uparrow$ \\
        Pipeline Parallelism & PP & - & $\downarrow$ & $\uparrow$ \\
        Expert Parallelism & EP & $\downarrow$ & $\downarrow$ & $\uparrow$ \\
        Data Parallelism & DP & $\uparrow$ & - & $\uparrow$ \\ 
        Fully-Sharded Data Parallel & FSDP & $\downarrow$ & $\downarrow$ & $\uparrow$$\uparrow$ \\
        Activation Recomputation & act & $\downarrow$ & $\downarrow$ & - \\
        Compute-Comm. Overlap & cc & $\uparrow$ & - & $\downarrow$ \\
        \bottomrule
    \end{tabular}
    }
    \vspace{-2ex}
    \label{tab:parallelism-techniques}
\end{table}

All models, except MoEs, use distributed optimizers like ZeRO-1~\cite{rajbhandari2020zero}, which partitions optimizer states across data-parallel ranks to reduce memory overhead. MoE models use standard optimizer configurations due to limitations in NeMo and Megatron-LM. 
We also evaluate optimization techniques, such as activation recomputation and compute-communication overlap, which affect efficiency based on model scale and system characteristics.
Note that both NeMo and Megatron-LM maps parallelism in the following order: TP $\rightarrow$ EP $\rightarrow$ DP $\rightarrow$ PP~\cite{shoeybi2019megatron,megatrondocs,nemoparallelism} 
This order determines how ranks are assigned and how communication occurs across devices, particularly in multi-node settings.
Table~\ref{tab:parallelism-techniques} summarizes the parallelism and optimization configurations in our study.

\begin{figure*}
    \centering
            \includegraphics[width=0.9\linewidth]{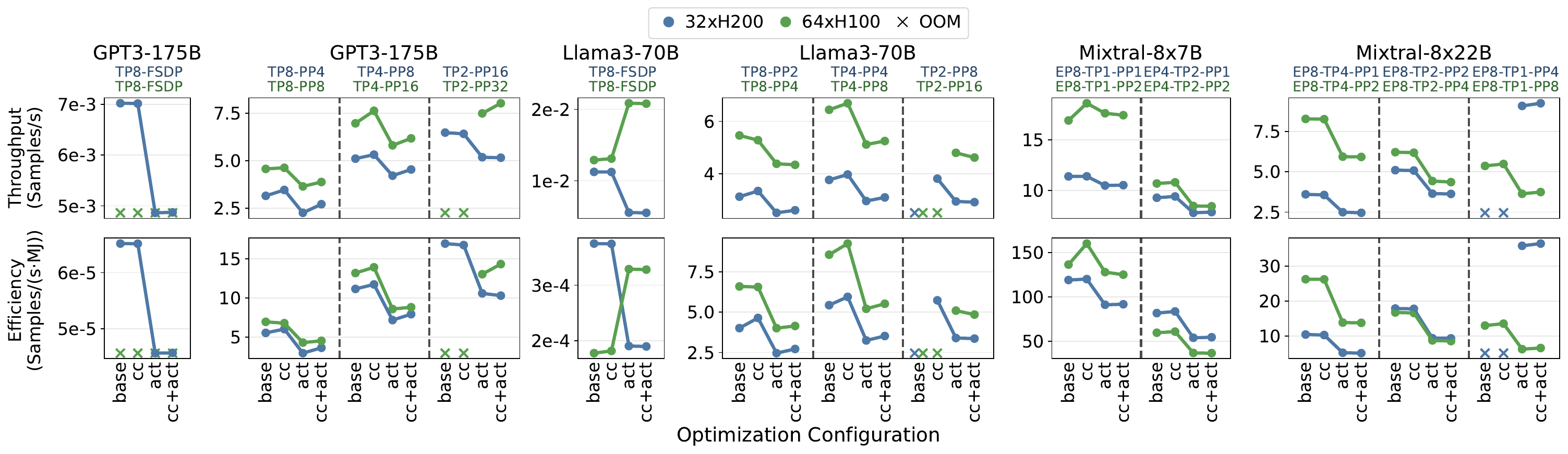}
    \vspace{-1em}
    \caption{Training throughput (top) and energy efficiency (bottom) for 64$\times$H100 vs. 32$\times$H200 clusters across parallelism and optimization settings. H100 outperforms under naïve setups, but H200 matches or exceeds it with optimized configurations.}
    \label{fig:efficiency-comparison}
\end{figure*}

\niparagraph{Terminology used across the paper.}
To characterize, we tune various options for configurations and use the following across the sections for uniformity. 
\textbf{``Base''} refers to the baseline setup, \textbf{``cc''} refers to compute-communication overlap, and \textbf{``act''} refers to activation recomputation.
Each distributed training strategy is shown as, \textbf{EP<e>-TP<t>-PP<p>}, for expert parallel width, tensor parallel width, and pipeline parallel depth.
If the number of GPUs exceeds the required model-parallel setup, we apply data parallelism across the remaining devices. 
For example, in a 32-GPU system, TP4-PP4 implies an additional DP of 2.
Similarly, \textbf{TP8-FSDP} indicates tensor parallelism across 8 GPUs, with 4 additional FSDP applied.
Finally, we use \textbf{``scale-up''} to denote systems with fewer but higher-capacity accelerators, and \textbf{``scale-out''} to refer to systems with more but lower-capacity accelerators.

\begin{table}
    \centering
    \caption{Hardware specifications of evaluated GPU clusters. 
    }
    \vspace{-0.7em}
    \label{tab:cluster-specs}
    \resizebox{1.0\columnwidth}{!}{
    \begin{tabular}{lccc}
        \toprule
        \textbf{Specification} & \textbf{HGX H200} & \textbf{HGX H100} & \textbf{MI250} \\
        \midrule
        \textbf{GPU Model} & NVIDIA H200 & NVIDIA H100 & AMD MI250 \\
        \textbf{GPU Architecture} & Hopper & Hopper & CDNA2 \\
        \textbf{Memory per GPU} & 141 GB HBM3 & 80 GB HBM3 & 64 GB HBM2e $\times$2 \\
        \textbf{Peak FP16/BF16} & 1.0 PFLOPS & 1.0 PFLOPS & 0.36 PFLOPS $\times$2 \\
        \textbf{GPUs per Node} & 8 & 8 & 8 (4$\times$2 GCDs) \\
        \textbf{Number of Nodes} & 4 & 8 & 4 \\
        \textbf{Intra-node Interconnect} & NVLink & NVLink & xGMI \\
        \textbf{Inter-node Interconnect} & 100 Gbps IB & 100 Gbps IB & 100 Gbps IB \\
        \textbf{GPU TDP} & 700 W & 700 W & 500 W \\
        \bottomrule
    \end{tabular}
    }
\end{table}

\subsection{Cluster Configuration}
Experiments are conducted on 3 GPU clusters with distinct hardware configurations, summarized in Table~\ref{tab:cluster-specs}.

\niparagraph{NVIDIA Cluster.} Training on NVIDIA platforms is performed using NeMo 24.07~\cite{kuchaiev2019nemo} with the Megatron-Core backend~\cite{shoeybi2019megatron}, built on CUDA 12.4, cuDNN 9.1.0, and NCCL 2.21.5.
We monitor GPU utilization, temperature, and power draw using the NVIDIA Management Library (NVML)~\cite{nvidia_nvml} integrated with Zeus for high-resolution profiling.
We evaluate two NVIDIA-based clusters with similar total memory capacity but differing compute density and communication characteristics: (1) an HGX H200 cluster with 4 nodes and (2) an HGX H100 cluster with 8 nodes.
While the H100 cluster provides higher aggregate compute, it incurs higher inter-node communication overhead due to the larger number of nodes.
By comparing these systems, we isolate the trade-offs between compute capacity, intra-node efficiency, and inter-node communication in real-world deployments.
The H200 cluster emphasizes tightly coupled communication within nodes, whereas the H100 cluster demonstrates how scaling across nodes affects synchronization and power behavior.

\niparagraph{AMD Cluster.} 
On AMD hardware, we use a ROCm-compatible Megatron-LM~\cite{shoeybi2019megatron} with ROCm 6.3 and RCCL 2.21 for collectives. 
Efficient kernels are supported through adapted forks of Apex, TransformerEngine, and FlashAttention.
Power and utilization metrics are collected using AMD-SMI~\cite{amd_smi}, extended via Zeus for fine-grained profiling.
The cluster consists of 4 nodes, each with 4$\times$MI250 accelerators. Each MI250 includes 2 GCDs, yielding 8 logical GPUs per node. Intra-node communication leverages xGMI, while nodes are connected via 100 Gbps InfiniBand.
We include the MI250 system to study performance and efficiency implications of chiplet-based GPU architectures, which introduce intra-package and inter-GCD communication asymmetries absent in NVIDIA's monolithic designs.
These differences enable us to examine their impact on parallelism strategies, resource utilization, and thermal behavior at scale.
Due to limited availability and lower memory capacity on AMD platforms, we evaluate them using smaller versions of GPT-3 and Llama-3.
Specifically, we scale down the models to fit within 4$\times$MI250 nodes, maintaining proportional relationships among key architectural parameters such as the number of layers, attention heads, and hidden dimensions.
The resulting models contain approximately 30 billion parameters.
Nonetheless, these runs provide valuable cross-platform insights that complement our broader analysis.

\begin{figure}
    \centering
            \includegraphics[width=1\linewidth]{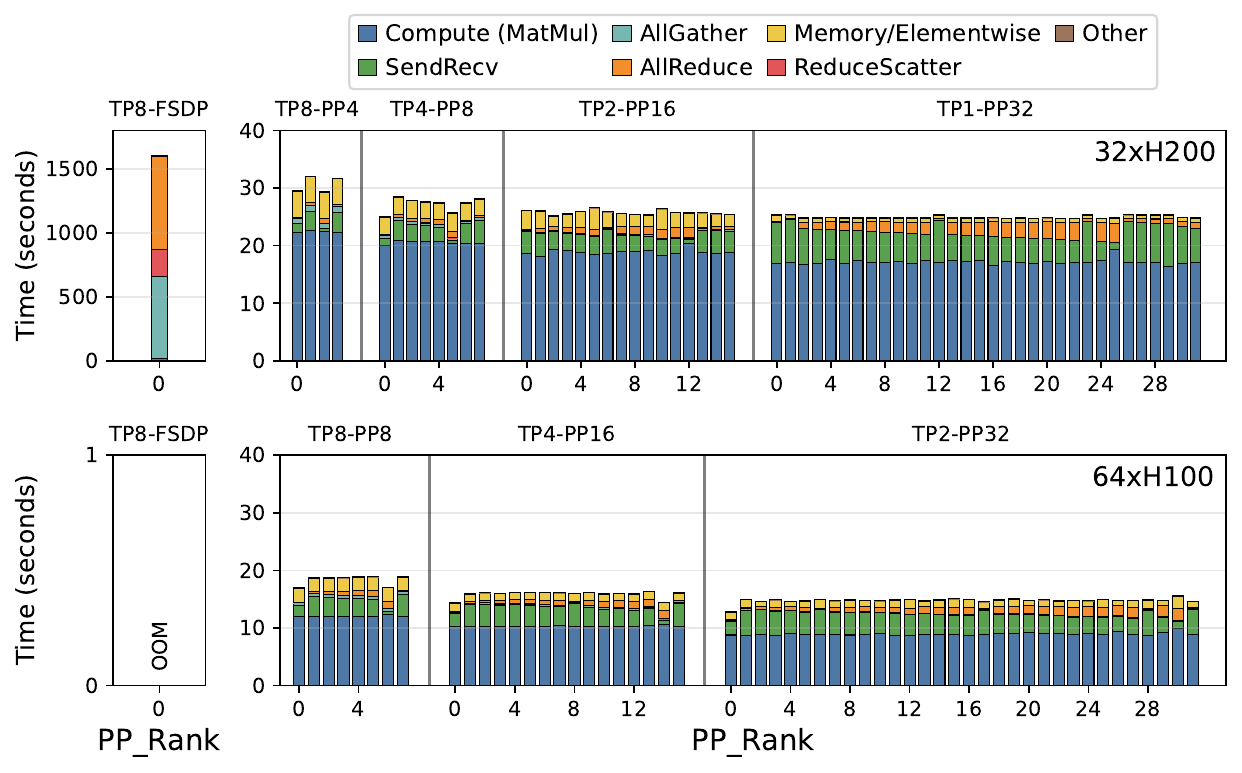}
    \vspace{-2em}
    \caption{Time across kernels for GPT3-175B training with all optimizations enabled across 32$\times$H200 and 64$\times$H100.}
    \label{fig:rank-distribution-compare}
    \vspace{-3ex}
\end{figure}

\section{Characterizing Scalable Distributed Training}\label{sec:characterization}

\subsection{Scale-up vs. Scale-out: Topology and Parallelism Co-Design Drive Efficiency}\label{subsec:scaling}

Figure~\ref{fig:efficiency-comparison} compares training throughput and energy efficiency between two GPU cluster configurations: 64$\times$H100 (scale-out) and 32$\times$H200 (scale-up).
We find that no single factor dictates the better scaling strategy; performance depends on model size, sparsity, and parallelism strategy, highlighting the importance of joint model-system co-design.

Overall, H100 scale-out outperforms H200 scale-up for smaller, compute-bound models like Llama3-70B and Mixtral-8x7B.
These models are predominantly compute-bound and therefore benefit from the H100 cluster's higher aggregate compute capacity (64$\times$H100 vs. 32$\times$H200 devices). As shown in Figure~\ref{fig:rank-distribution-compare}, H100 spends less time on compute across all parallelism schemes, despite H200's larger memory and denser topology.
However, for larger or sparser, communication-bound models like GPT3-175B and Mixtral-8x22B, the performance gap narrows -- or even reverses. In GPT3-175B and Mixtral-8x22B, we observe higher communication volume due to model and expert parallelism. Here, H200’s 1.76$\times$ larger memory and smaller network enable better communication locality and reduced inter-node traffic. This allows H200 to match or surpass H100 performance in certain configurations, such as Mixtral-8x22B under EP8–TP1–PP4, where expert and tensor parallelism communication are largely confined within nodes.

\begin{figure}
    \centering
        \includegraphics[width=0.9\linewidth]{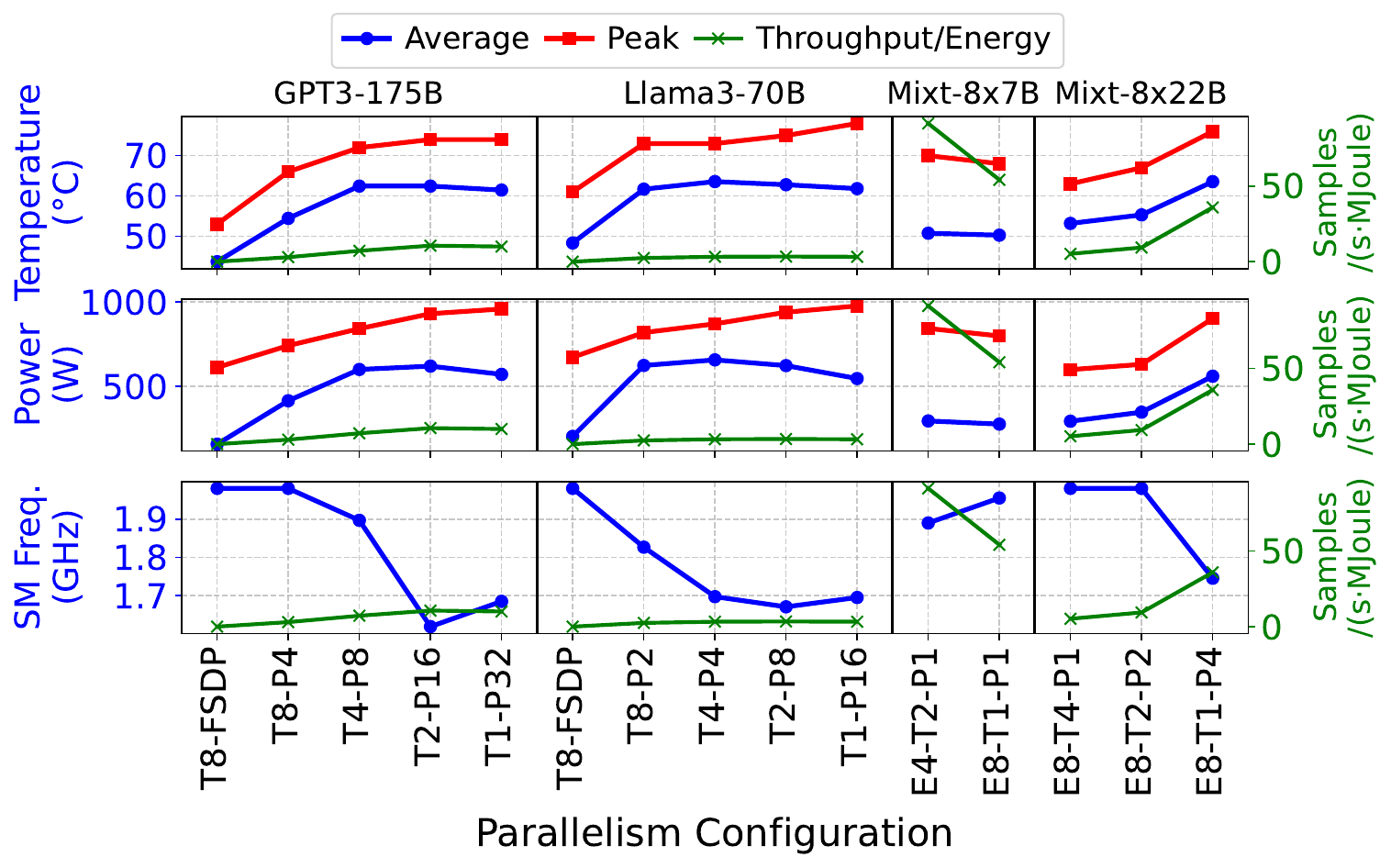}
        \includegraphics[width=0.9\linewidth]{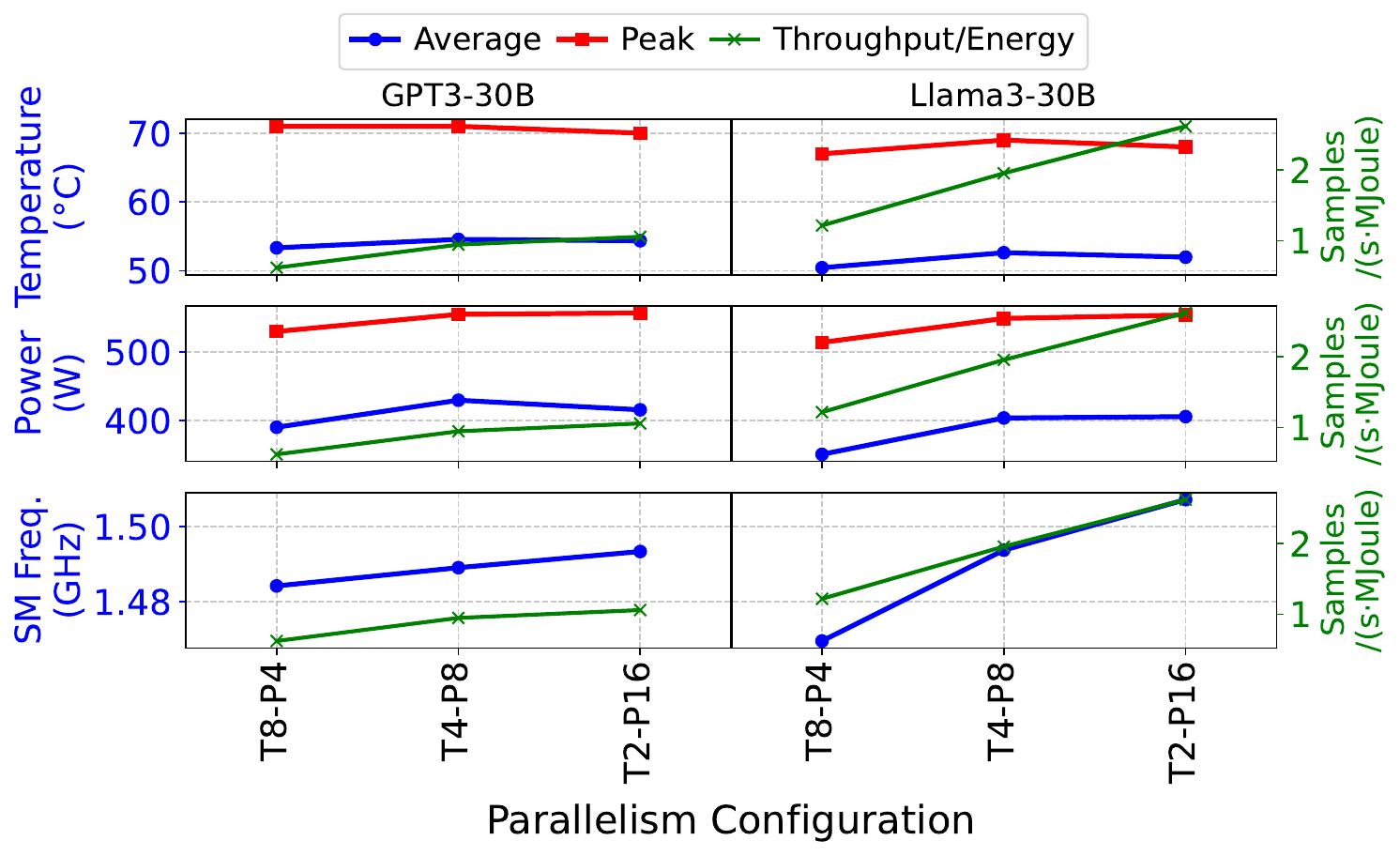}
    \vspace{-1em}
    \caption{GPU temperature, power, and frequency for H200 (top) and MI250 (bottom) clusters across different models and parallelism strategies, with activation recomputation enabled for additional configurations.}
    \label{fig:parallel-sweep}
    \vspace{-3ex}
\end{figure}

Interestingly, the energy efficiency trends mirror this crossover behavior. While H100 generally trains faster and thus consumes less total energy, the H200 cluster exhibits superior efficiency under communication-heavy settings. For example, in GPT3-175B with TP2–PP16, and Mixtral-8x22B across multiple configurations, H200 outperforms H100 in both throughput and energy per token. This advantage is amplified by H200’s ability to achieve comparable performance using only half the number of GPUs.

These findings yield our first major insight: the effectiveness of scale-up versus scale-out hardware cannot be determined in isolation from model characteristics and parallelism strategies. Small, compute-bound models (dense or sparse) benefit from scale-out systems with higher aggregate compute, while large or communication-heavy models (especially sparse ones with expert parallelism) are better suited for scale-up architectures that reduce communication costs through node-locality.

That said, while scale-up shows promise for energy-efficient training, its success hinges on carefully designed parallelism and hardware-aware communication strategies. Simply deploying denser hardware is insufficient, efficient execution requires tightly integrated software-hardware co-design to align the parallelism structure with system topology. We further explore the impact of parallelism configuration in Section~\ref{subsec:parallelism}.

\vspace{0.5em}
\noindent\fbox{%
    \parbox{\columnwidth}{
    \textit{\textbf{Insight: Hardware potential alone does not guarantee efficiency.}
    Despite H200’s higher GPU density and larger memory capacity, its full potential is difficult to realize without hardware-aware system design. Naively selected parallelism strategies can negate its density advantage due to pipeline stalls and communication bottlenecks. However, when communication-intensive workloads are effectively confined within fewer nodes, the 32$\times$H200 cluster can match or even outperform the 64$\times$H100 cluster in both performance and energy efficiency. This highlights the importance of co-designing model parallelism strategies in alignment with cluster topology.
    }}
}
\subsection{System-Level Impact of Parallelism Choices}
~\label{subsec:parallelism}

To better understand how hardware topology and model characteristics interact, we evaluate a range of parallelism configurations and optimization techniques across diverse models and hardware setups.
Figure~\ref{fig:parallel-sweep} summarizes power, temperature, and frequency metrics for H100 and MI250 GPUs across various parallelism configurations, alongside corresponding throughput.
Deeper pipeline parallelism (e.g., PP16–32) generally improves GPU utilization and reduces communication overhead (due to its simpler send and receive communication patterns), yielding higher runtime efficiency. 
However, this comes at the cost of significantly higher peak power and thermal load, particularly for larger models like GPT3-175B.
However, this comes with higher peak power and thermal imbalance, particularly in large models like GPT3-175B. Pipelining too deeply (e.g., TP1–PP32) can introduce diminishing returns or even degrade performance due to increased pipeline bubbles and reduced per-stage granularity. Moderately deep pipelining (e.g., TP2–PP16) achieves a better tradeoff, sustaining high utilization while keeping thermal imbalance manageable.

\begin{figure}
    \centering
    \begin{subfigure}[]{\linewidth}
        \centering
        \includegraphics[width=0.95\linewidth]{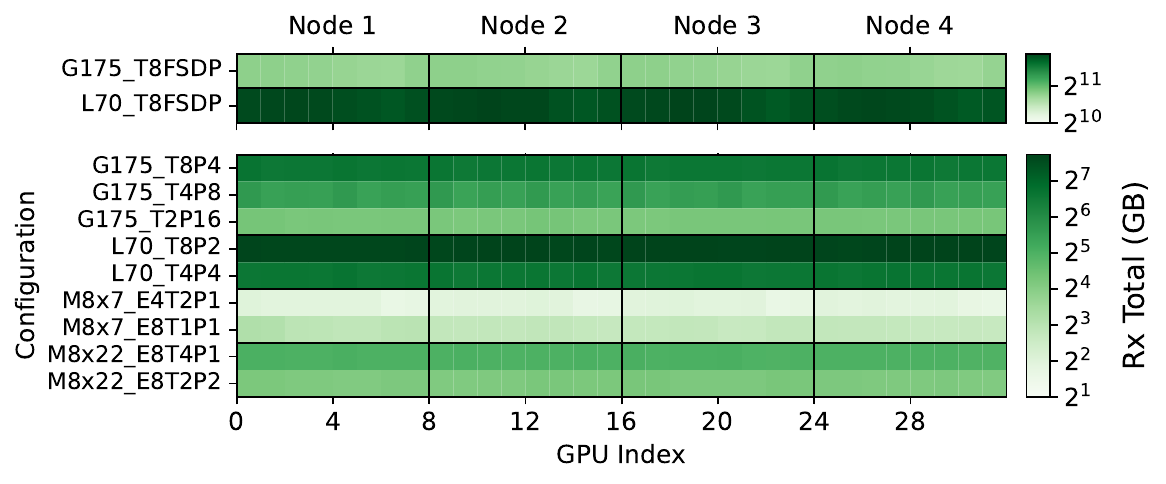}
        \vspace{-0.5em}
        \caption{Total NVLink Rx traffic heatmap.}
        \label{fig:nvlink-rx}
    \end{subfigure}
    \vfill
    \begin{subfigure}[]{\linewidth}
        \centering
        \includegraphics[width=0.95\linewidth]{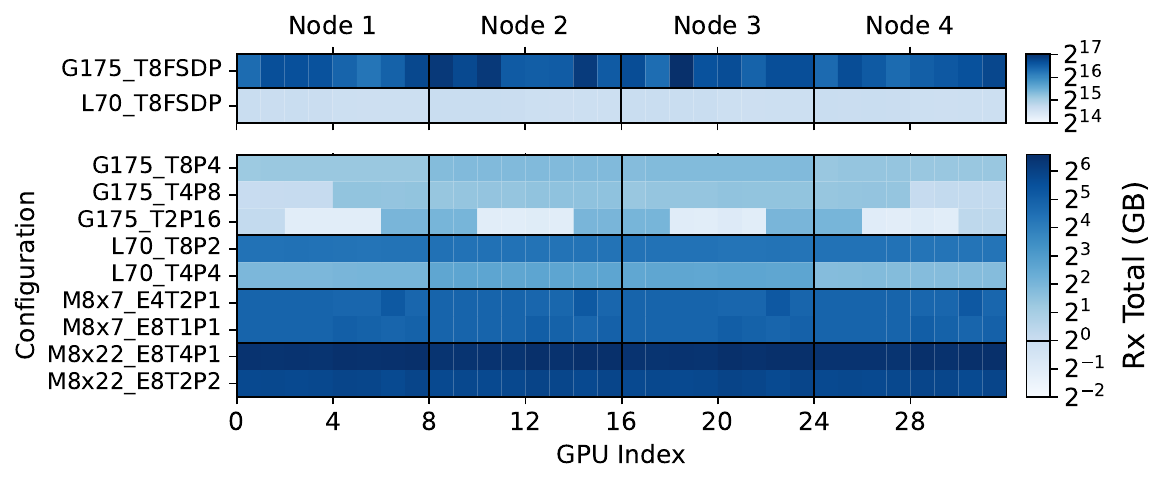}
        \vspace{-0.5em}
        \caption{Total PCIe Rx traffic heatmap.}
        \label{fig:pcie-rx}
    \end{subfigure}
    \vspace{-1em}
    \caption{Per-GPU total NVLink and PCIe traffic distribution of the HGX H200 cluster during model training.}
    \label{fig:parallel-traffic}
    \vspace{-1em}
\end{figure}

In contrast, TP-heavy configurations draw less power but incur significantly higher communication overhead. 
As TP depth increases--especially in models like Mixtral--we observe a sharp rise in PCIe and NVLink traffic, exceeding 70GB in some cases. This is due to intensified inter-node communication, particularly when tensor and expert parallelism span multiple nodes. Figure~\ref{fig:parallel-traffic} shows this overhead, highlighting that TP-heavy strategies amplify all-to-all communication patterns, especially detrimental under sparse expert routing.
Using TP with FSDP introduces additional inter-node communication, as optimizer states, weights, and gradients are sharded across nodes.
As a result, the TP8-FSDP configuration shows some degradation in performance and energy efficiency compared to TP-PP setups.

\begin{figure}
    \centering
    \includegraphics[width=0.85\linewidth]{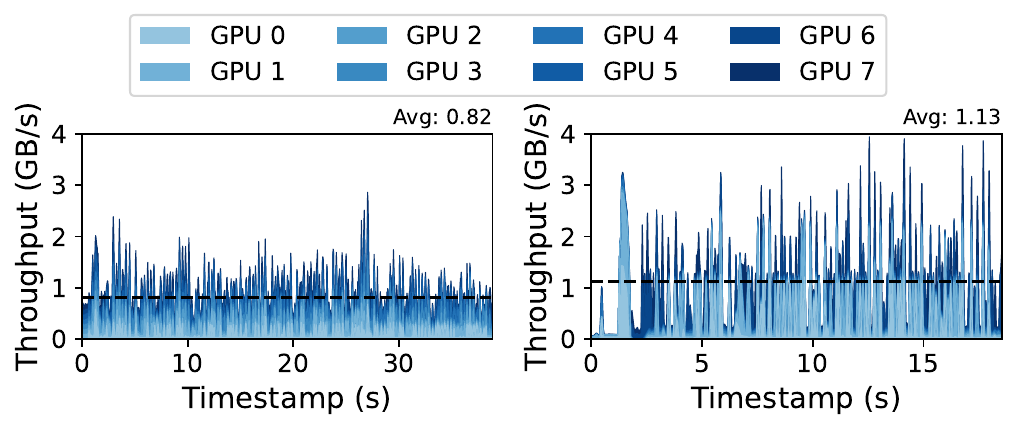}
    \vspace{-1.5em}
    \caption{Aggregate PCIe throughput over time across 8 GPUs in an H200 node during GPT3-175B training with TP8-PP4 (left) and TP2-PP16 (right).}
    \label{fig:pcie-stacked}
    \vspace{-2ex}
\end{figure}

\begin{figure}
    \centering
            \includegraphics[width=1\linewidth]{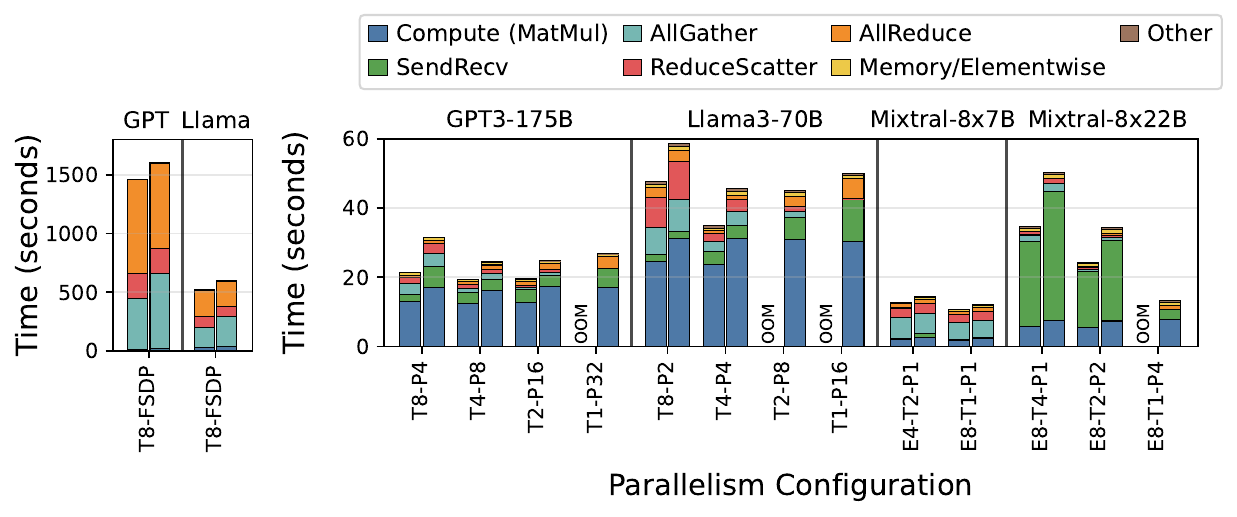}
        \label{fig:parallel-distribution-combined}
    \vspace{-2.5em}
    \caption{
    Breakdown of latency by kernel. Kernel times are averaged across all GPU ranks. 
    Within each parallelism configuration, each stacked bar represents results 
    without (left) and with (right) activation recomputation.
    }
    \label{fig:parallel-distribution}
    \vspace{-2ex}
\end{figure}

PP-heavy configurations show lower PCIe traffic, with communication more concentrated across boundary GPUs. This concentrated pattern improves bandwidth utilization by transferring larger data chunks over fewer endpoints, thereby reducing per-link contention and communication overhead, as shown in Figure~\ref{fig:pcie-stacked}.

This behavior is further reflected in the kernel timing breakdown (Figure~\ref{fig:parallel-distribution}). Dense models spend over 50\% of kernel time on computation, while sparse MoE models are dominated by communication. For example, in Mixtral-8x22B, SendRecv time drops sharply when TP width is reduced -- despite unchanged EP degree -- because communication becomes more localized, avoiding inter-node transfers and reducing overhead.
Each TP group spans more GPUs within a node, limiting the intra-node placement of other parallelism groups like EP and forcing inter-node communication for all-to-all.
Reducing TP width allows more GPUs within a node to be dedicated to EP, thus all-to-all remains within a node, significantly lowering inter-node traffic and improving communication efficiency.

Our results expose key inefficiency when combining TP and PP; TP+PP configurations trigger sparse, uncoordinated SendRecv calls across GPUs, which fail to leverage data chunking or collective scheduling. These patterns underutilize available PCIe and network bandwidth, increasing communication latency and stalling compute stages. Figure~\ref{fig:rank-distribution-compare} shows this effect clearly--communication time skews heavily across ranks in TP8–PP4 due to PCIe and NIC contention. Unlike NVIDIA DGX systems that assume dedicated paths, scale-out clusters share communication interfaces, leading to contention and bottlenecks.

These findings expose a critical gap in current ML systems design: parallelism strategies are rarely co-designed with awareness of real-world hardware variability. Existing frameworks prioritize algorithmic metrics like FLOPs or ideal scaling but ignore practical factors like interconnect congestion or topology asymmetry. 
For example, EP configurations suffer from large numbers of fine-grained SendRecv kernels, introducing irregular communication with low payload sizes. Without proper batching or locality-aware routing, GPUs compete for limited PCIe lanes and NIC bandwidth, degrading efficiency.
In contrast, PP-heavy setups -- due to their P2P-style communication and stage isolation -- show much more stable communication profiles and balanced rank-level kernel time. As shown in Figure~\ref{fig:pcie-stacked}, PP-heavy configurations transfer larger data chunks over fewer paths, reducing per-link contention and improving effective bandwidth utilization.
In summary, as LLM training increasingly pushes infrastructures to their physical and thermal limits, this gap becomes a central bottleneck.

\begin{figure}
\centering
\includegraphics[width=0.85\linewidth]{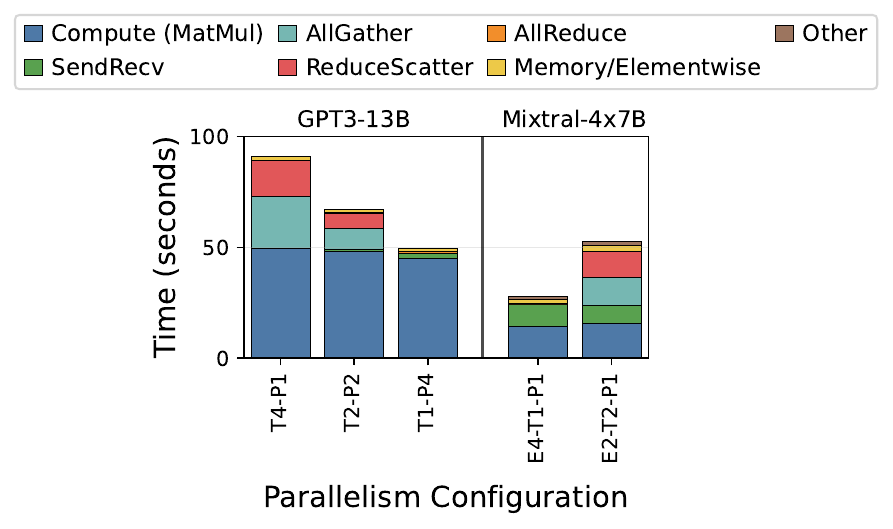}
\vspace{-1.3em}
\caption{Kernel latency breakdown for 1-GPU-per-node setup: PP-heavy regions reduce communication time, while TP-heavy regions are bottlenecked by network bandwidth.}
\label{fig:1gpu4node-comm}
\vspace{-1em}
\end{figure}

To validate the generality of our findings under more balanced interconnect conditions, we ran experiments using a single-GPU-per-node setup across four nodes.
This configuration reduces PCIe and NIC contention, improving inter-node bandwidth and creating a more uniform communication topology.
Due to HBM limits, we used smaller models, GPT3-13B~\cite{gpt-3} and Mixtral-4x7B (a reduced version of Mixtral-8x7B).
As shown in Figure~\ref{fig:1gpu4node-comm}, communication time drops significantly for PP-heavy workloads. TP-heavy setups, however, still show over 10$\times$ higher communication time than PP-only configurations. 
Mixtral also faces significant bottlenecks, with communication exceeding 50\% of total latency.
These findings reinforce the importance of topology-aware parallelism strategies, even in more balanced network environments.

\begin{figure*}
    \centering
        \includegraphics[width=0.95\linewidth]{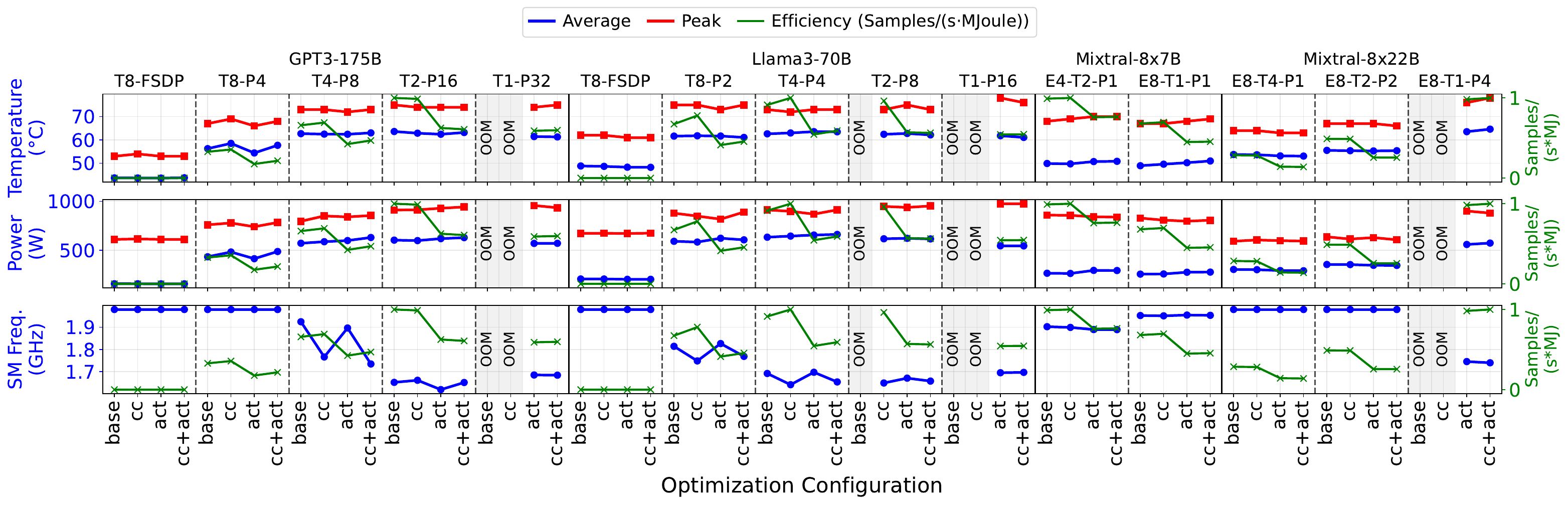}
    \label{fig:optim-sweep}
    \vspace{-1em}
    \caption{GPU power, temperature, and clock frequency on the H200 cluster across different models, parallelism configurations, and optimization techniques, highlighting system-level trade-offs. Efficiency values are normalized per model, relative to that model’s most efficient configuration.}
    \label{fig:optim-sweep}
    \vspace{-1em}
\end{figure*}

\vspace{0.5em}
\noindent\fbox{%
\parbox{\columnwidth}{%
\textit{\textbf{Insight: Communication Variability shapes Load Balance and System Reliability.} \
If communication collectives do not account for the underlying physical topology, TP- and EP-heavy configurations suffer from high communication overhead and increased variability due to cross-node traffic and load imbalance.
While PP-heavy strategies help reduce communication skew, they can create thermal and power hotspots. Achieving efficient and reliable training demands balanced parallelism and topology-aware collectives that localize communication wherever possible.
}}%
}

\begin{figure}[h]
    \centering
    \includegraphics[width=0.9\linewidth]{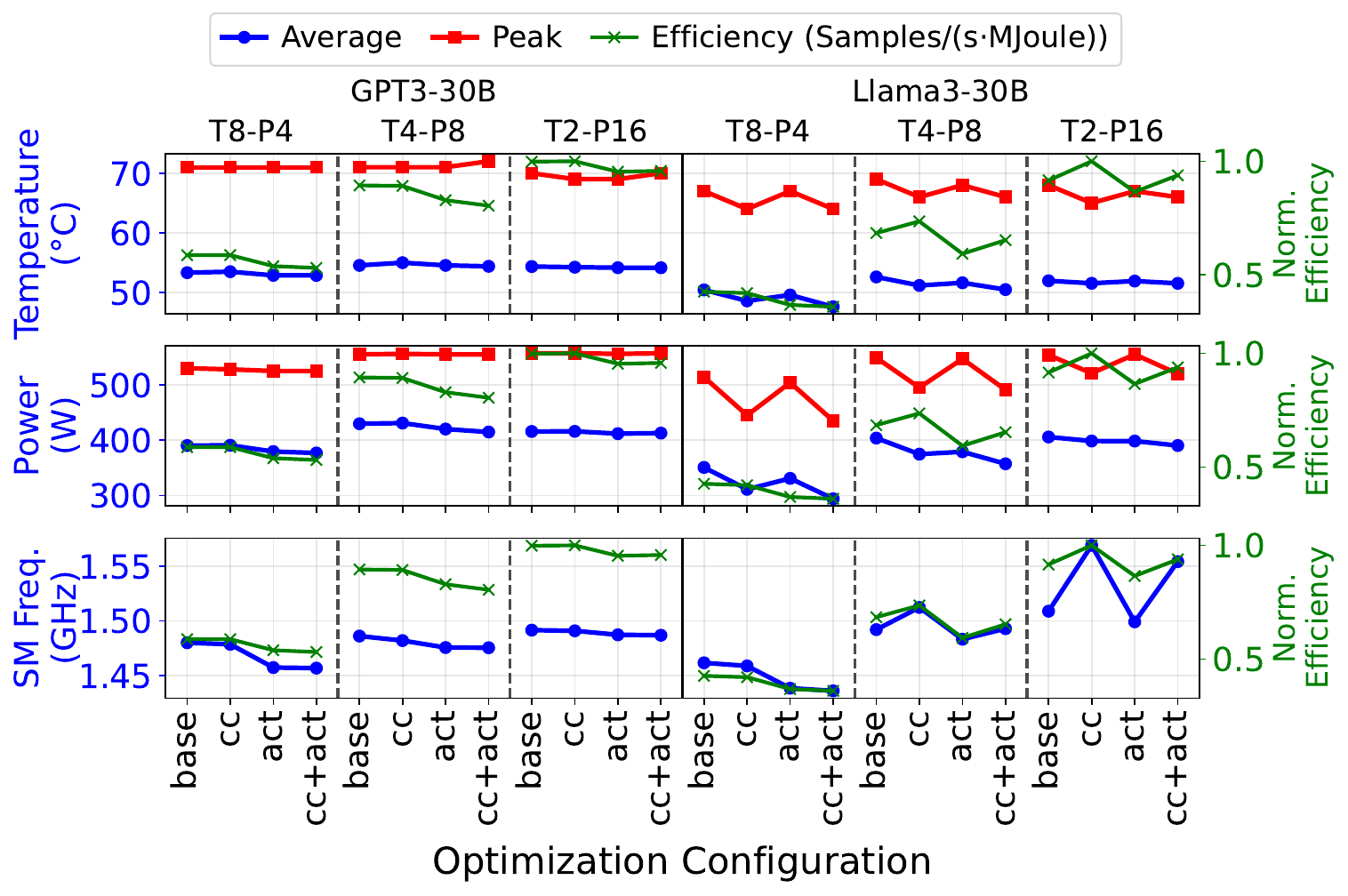}
    \label{fig:optim-sweep-amd}
    \vspace{-1em}
    \caption{GPU power, temperature, and clock frequency on the MI250 cluster. Efficiency values are normalized per model, relative to the best-performing configuration.
}
    \label{fig:optim-sweep-amd}
    \vspace{-1ex}
\end{figure}

\subsection{Effects of Training-Time Optimizations}\label{subsec:optimization}
To address emerging bottlenecks and maximize resource utilization, additional training-time optimizations are commonly employed. In this section, we evaluate three such techniques -- activation recomputation, compute-communication overlap, and low-rank adaptation -- and analyze their effects on system performance, load balance, and thermal behavior across various parallel configurations.
These techniques interact with parallelism strategies in complex ways, shaping trade-offs between throughput, energy efficiency, and system reliability. Our goal is to understand how they can be leveraged in practice to expand the design space and improve end-to-end training outcomes.

\niparagraph{Activation Recomputation.}
Activation recomputation~\cite{activationrecomputation} trades increased computation for reduced memory consumption by recomputing intermediate activations during backpropagation rather than storing them in memory. This approach is particularly useful when training large models, where memory capacity is often a limiting constraint.
As shown in Figure~\ref{fig:optim-sweep} and Figure~\ref{fig:optim-sweep-amd}, activation recomputation enables deeper pipeline-parallel (PP) configurations that are otherwise infeasible under stashing, where activations are stored. However, this flexibility comes at a cost: training efficiency consistently drops across all tested models and hardware setups due to the extra computation required. We also observe a small increase in clock throttling, likely driven by the prolonged high-power execution phases introduced by recomputation.
The negative impact on efficiency is especially evident in deep PP configurations, where recomputation lengthens forward and backward stages and exacerbates stage-level imbalances. Figure~\ref{fig:parallel-distribution} further confirms a shift in kernel latency distributions, showing increased compute times across configurations and a higher total kernel time.

Despite this trade-off, activation recomputation unlocks configurations that were previously infeasible, for instance, the E8-T1-P4 setup on Mixtral-8x22B. It also benefited from reduced PCIe and all-to-all communication, as expert parallelism became fully localized within each node. As seen in Figure~\ref{fig:optim-sweep}, this setup achieved over 2$\times$ the training efficiency of the best-performing baseline configuration.
\textit{This result highlights how activation recomputation can expand the design space and enable better trade-offs between compute, communication, and memory. While the technique imposes a computational overhead, the ability to unlock more efficient configurations offsets these costs in some cases, especially for memory-bound models or parallelism layouts.}

\niparagraph{Compute-Communication Overlap.}
Compute-communication overlap (CC-overlap) aims to hide communication latency by overlapping it with concurrent compute operations. This is typically achieved by replacing large, monolithic communication kernels with finer-grained memory or element-wise operations that can be interleaved with computation. CC-overlap is effective in communication heavy configurations, where idle time due to data movement can become a bottleneck.

Figures~\ref{fig:optim-sweep} and~\ref{fig:optim-sweep-amd} show the system-level impact of CC-overlap across models and configurations, capturing GPU power and temperature. The benefits vary by parallelism layout -- TP-heavy configurations with high communication overheads see the most improvements, due to reduced communication time and better resource utilization.
Llama3-70B, trained using data parallelism (DP) with a distributed optimizer, shows strong gains from CC-overlap. The distributed optimizer provides more compute work per rank, enabling higher degrees of overlapping with communication. In contrast, PP-heavy configurations (e.g., GPT-3 with T2-P16) often experience reduced benefits or even performance degradation, largely due to thermal throttling. In these setups, concurrent compute and communication place greater stress on the memory and SM subsystems, leading to overheating and eventual slowdowns.
Similarly, in TP8–FSDP, while the proportion of overlapped compute is substantial, it has a negligible impact on end-to-end step time. 
Communication latency dominates runtime, so overlap improves GPU utilization but doesn’t resolve the main bottleneck—limited network bandwidth.

\begin{figure}
    \centering
            \includegraphics[width=1\linewidth]{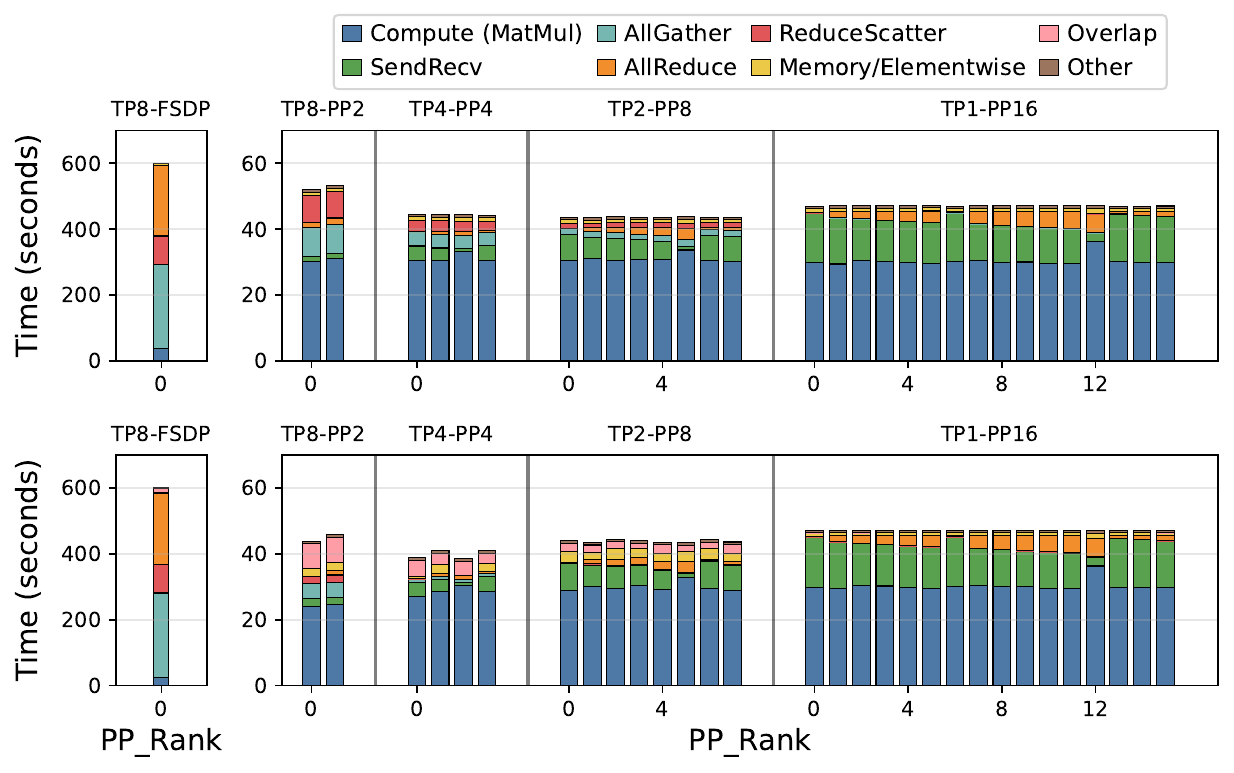}
        \label{fig:rank-distribution}
    \vspace{-2em}
    \caption{Breakdown of latency by kernel for Llama3-70B training across pipeline-parallel ranks, without overlap (top) and with CC-overlap (bottom).}
    \label{fig:rank-distribution-ccoverlap}
    \vspace{-3ex}
\end{figure}

\begin{figure}
    \centering
        \includegraphics[width=0.9\linewidth]{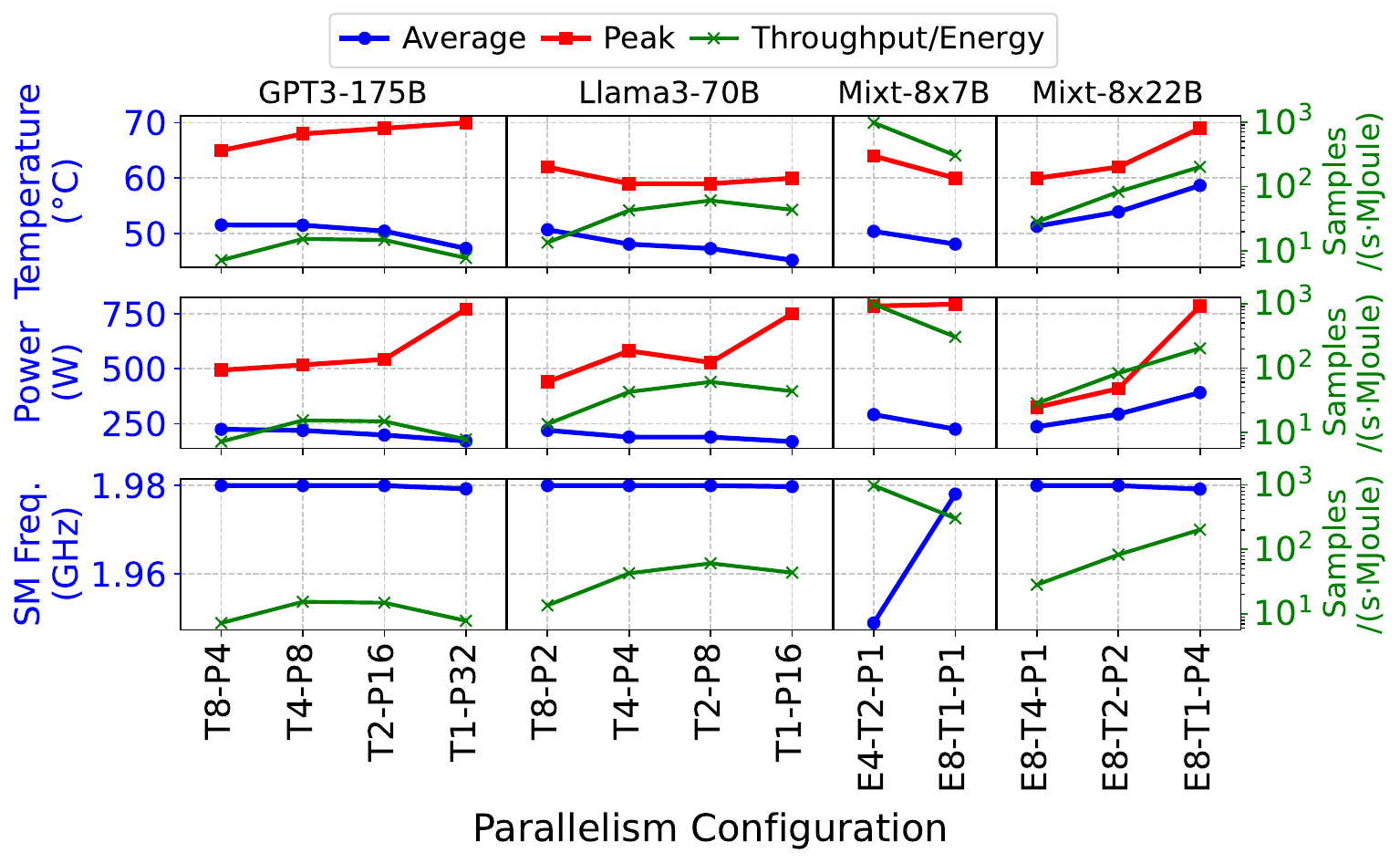}
    \vspace{-1em}
    \caption{GPU temperature, power, and frequency on the H200 cluster during LoRA fine-tuning.}
    \label{fig:lora-efficiency}
    \vspace{-1em}
\end{figure}

\begin{figure*}
    \centering
        \includegraphics[width=0.9\linewidth]{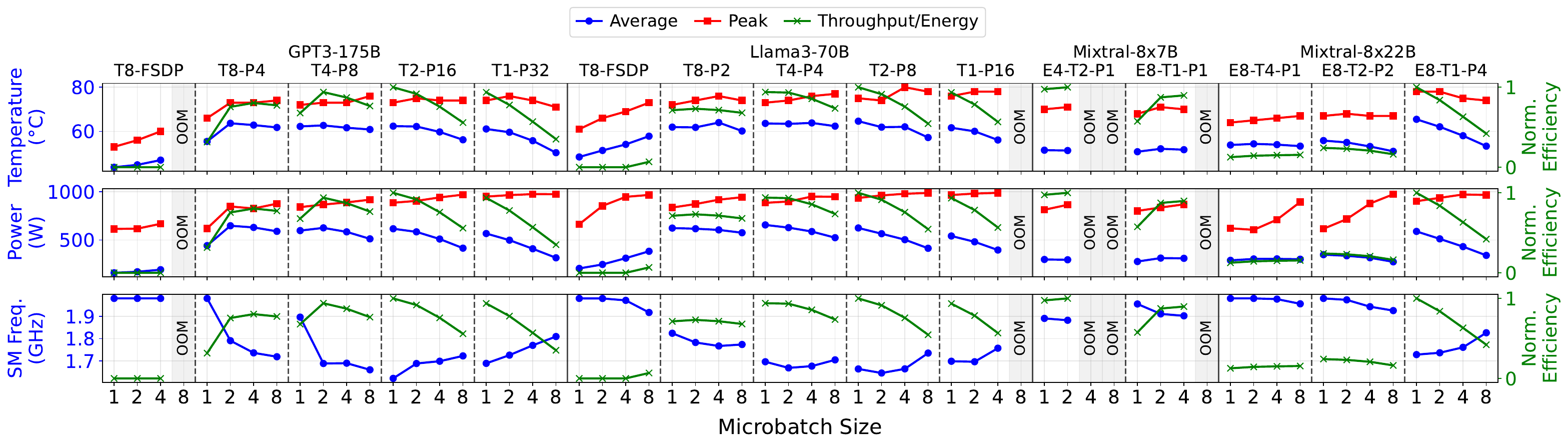}
    \label{fig:ubatch-sweep}
    \vspace{-1em}
    \caption{GPU power, temperature, and clock frequency on the H200 cluster across models, parallelism configs, and microbatch sizes. Activation recomputation enabled. Efficiency normalized per model (best = 1).}
    \label{fig:ubatch-sweep}
    \vspace{-1em}
\end{figure*}

\begin{figure}
    \centering
    \includegraphics[width=0.95\linewidth]{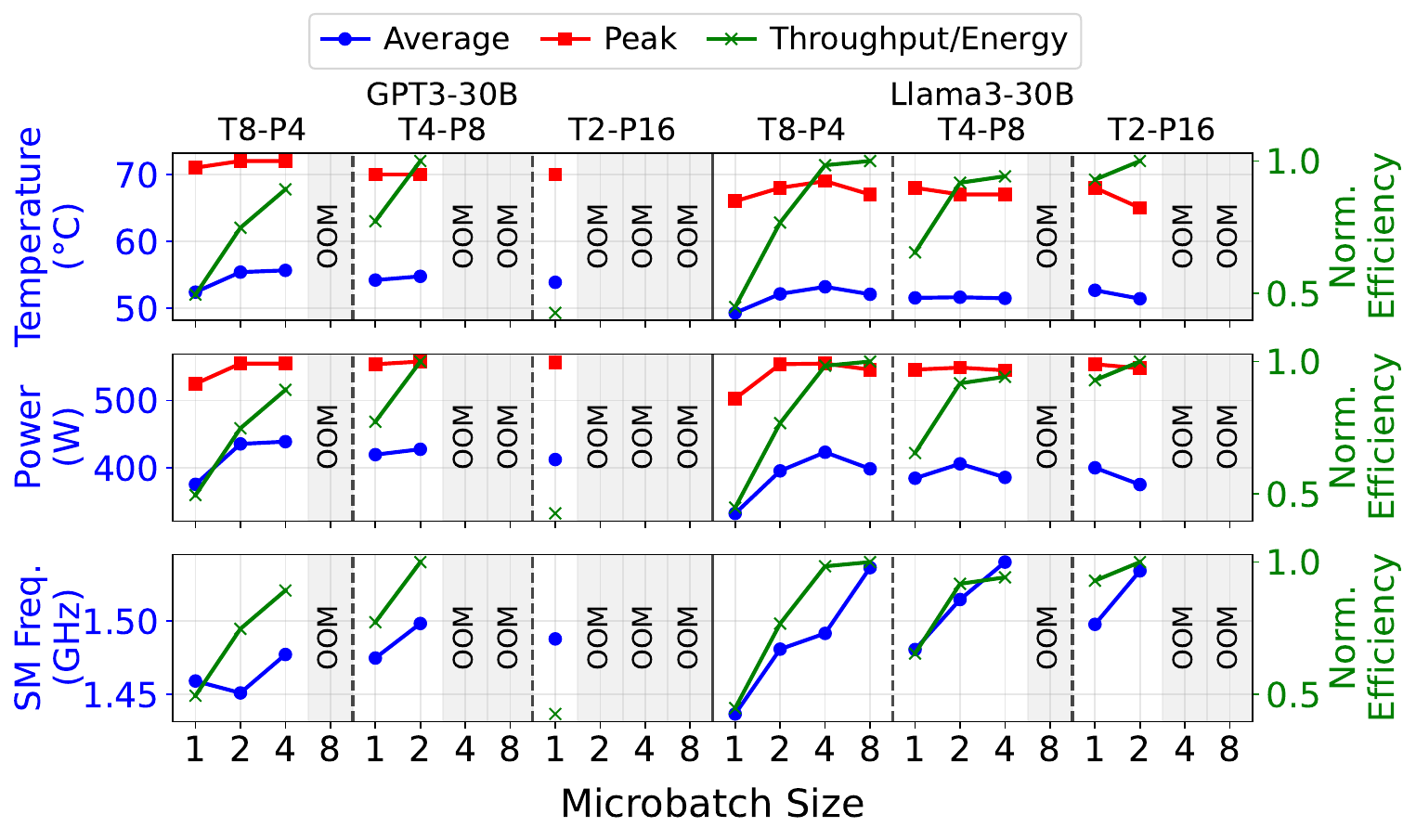}
    \label{fig:ubatch-sweep-amd}
    \vspace{-1em}
    \caption{GPU power, temperature, and clock frequency on the MI250 cluster with activation recomputation enabled. Efficiency numbers are normalized per model. Efficiency normalized per model (best = 1).}
    \label{fig:ubatch-sweep-amd}
    \vspace{-2.5ex}
\end{figure}

Across configurations, CC-overlap consistently increases peak temperature, as shown in Figure~\ref{fig:optim-sweep}. This rise in thermal stress can negate the performance gains if not properly managed, introducing a trade-off between utilization and thermal reliability. The effectiveness of CC-overlap is not solely determined by how much overlap is achieved -- it also depends on the system’s ability to sustain high utilization without triggering throttling or introducing imbalance.
To illustrate this, we examine kernel execution profiles for Llama3 training across pipeline-parallel ranks in Figure~\ref{fig:rank-distribution-ccoverlap}. Compared to the non-overlapped baseline, CC-overlap replaces large communication kernels with finer operations that can be scheduled alongside compute kernels. However, we observe that the compute kernel durations also increase, especially in TP-heavy setups. This reflects heightened contention for shared resources and indicates that if overlap is not carefully tuned, the total execution time may increase rather than decrease.
\textit{In summary, CC-overlap is effective in reducing communication overhead and improving efficiency in communication-bound scenarios, but its full benefit depends on careful integration with parallelism strategy and thermal management to avoid resource contention and throttling-induced regressions.}

\niparagraph{LoRA Finetuning.}
To improve training efficiency for large-scale models, we evaluate parameter-efficient finetuning using low-rank adaptation (LoRA)~\cite{lora}.
LoRA inserts trainable low-rank adapters into attention and feedforward layers, drastically reducing the number of updated parameters while maintaining strong performance.
This approach lowers system requirements, making LoRA well-suited for finetuning massive models on bandwidth- or power-constrained infrastructure.
Figure~\ref{fig:lora-efficiency} presents power, temperature, and frequency metrics during LoRA fine-tuning.
LoRA achieves over 10$\times$ higher training efficiency compared to full model training, mainly due to fewer parameters and reduced gradient synchronization.
It also lowers GPU power and temperature, while showing a similar trend to pretraining across different parallelism strategies.
\textit{These results position LoRA as a practical, scalable solution for finetuning large models—boosting efficiency and system stability without architectural changes or hardware upgrades.}

\section{Limits of Microbatch Scaling: Higher Utilization vs. System Stress}\label{subsec:microbatch}

In at-scale training, increasing microbatch size is often assumed to be a straightforward way to improve performance particularly when memory capacity allows.
Larger microbatches reduce the frequency of inter-GPU communication, offer better compute-to-communication ratio, and generally push training closer to being compute-bound.
However, our analysis reveals that this intuition does not always hold: with certain GPUs, increasing microbatch size can degrade system efficiency, stress hardware, and reduce overall training throughput.

We analyze this effect across multiple models and parallelism configurations, as shown in Figures~\ref{fig:ubatch-sweep} and~\ref{fig:ubatch-sweep-amd}.
On NVIDIA clusters, larger microbatches improve performance in FSDP- and TP-dominated setups (e.g., TP8–FSDP, TP8–PP4 for GPT-3). However, efficiency drops sharply in pipeline-heavy configurations like TP2–PP16.
This behavior generalizes across models, reflecting that microbatch size interacts closely with parallel execution structure, communication overheads, and scheduling skew.
For the MI250 cluster, in contrast, increasing the microbatch size generally improves training efficiency. 
This is primarily because memory capacity is reached before the GPU experiences any thermal stress, i.e., thermal constraints do not limit performance. As microbatch size increases, the workload becomes more compute-intensive, prompting the GPU to boost its clock frequency. This leads to improved SM utilization and higher overall training efficiency.

\begin{figure}[]
    \centering
    \includegraphics[width=1\linewidth]{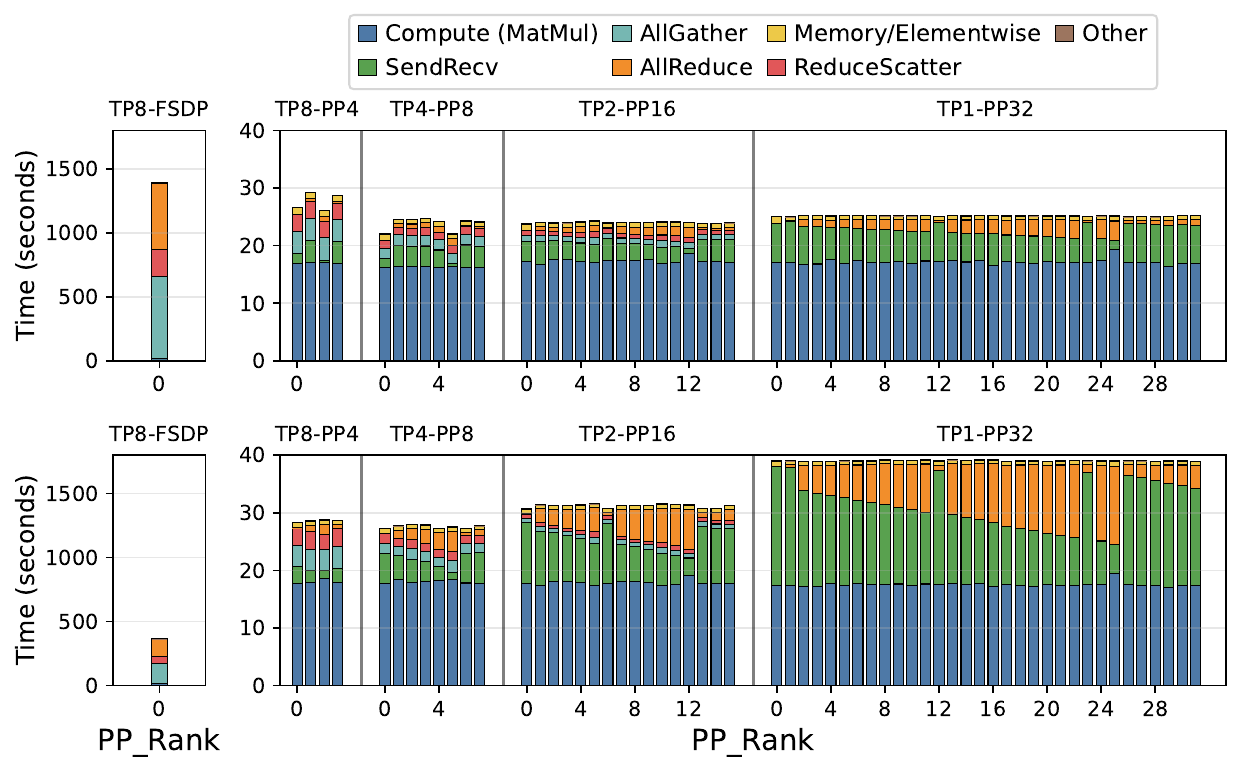}
    \label{fig:ubatch-distribution}
    \vspace{-1em}
    \caption{Breakdown of kernel latency for H200 cluster with microbatch 1 (top) and 4 (bottom).}
    \label{fig:ubatch-distribution}
    \vspace{-1em}
\end{figure}

Beyond performance, system-level metrics reveal that the benefits of larger microbatches come with significant costs, especially when the system is stressed.
Total power consumption and average GPU temperature tend to increase in line with efficiency gains for NVIDIA systems.
However, peak power and thermal levels rise consistently with larger microbatches -- regardless of whether training throughput improves.
This leads to two important system effects: in TP-heavy configurations, higher peak power triggers clock throttling, reducing compute efficiency; in PP-heavy setups (such as TP2–PP16 and TP1–PP32), we observe increased average clock frequency as pipeline stalls emerge, resulting in more bursty execution patterns that intermittently under-utilize compute resources.

To better understand the underlying performance characteristics, we analyze the per-rank breakdown of latency by kernel for GPT3-175B under microbatch sizes of 1 and 4, as shown in Figure~\ref{fig:ubatch-distribution}.
At smaller microbatch sizes, communication time dominates in TP-heavy setups, with significant skew across ranks.
This imbalance is alleviated by increasing PP depth -- but extreme pipeline configurations (e.g., TP1–PP32) reintroduce inefficiencies as communication costs rise again.
Larger microbatches improve execution uniformity across ranks by reducing skew, but at the cost of higher overall communication time and bandwidth contention, particularly in PP-heavy setups where SendRecv and AllReduce operations become bottlenecks.
On the other hand, in TP8–FSDP, increasing the microbatch size from 1 to 4 yields over a 3$\times$ speedup in step time.
This improvement is largely due to coarser-grained communication with larger microbatches, which better utilizes limited network bandwidth (Section~\ref{subsec:parallelism}).
As such, while increasing microbatch size can improve training performance under the right conditions, it is far from a universally effective knob.
Careful tuning is essential—not just for maximizing throughput, but also for avoiding system strain and performance reduction in configurations where hardware and communication limits become dominant.

\vspace{0.5em}
\noindent\fbox{%
    \parbox{\columnwidth}{%
       \textit{\textbf{Insight: Larger microbatch size is not a universally effective knob due to hardware and system limits.} 
        Larger microbatches are not always better. Once past an optimal point, further increases yield diminishing returns or harm performance. This is due to: (i) communication bandwidth saturation, (ii) pipeline-induced stalls and under-utilization, and (iii) increased power draw and thermal stress triggering clock throttling. These factors can offset efficiency gains due to larger microbatches.}}}

\section{Exposing the Cost of Uniform Optimizations under Thermal Imbalance}\label{sec:thermal-imbalance}

\begin{figure}
    \centering
    \begin{subfigure}[t]{0.48\linewidth}
    \centering
    \includegraphics[width=0.95\linewidth]{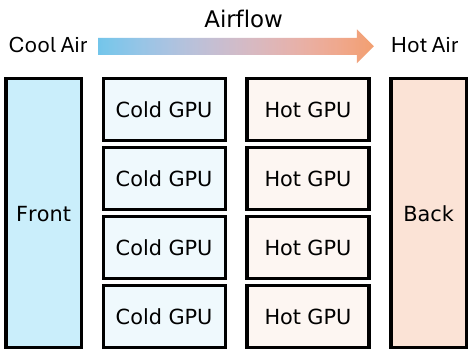}
    \caption{Airflow and cooling layout of the HGX node.}
    \label{fig:hgx}
    \end{subfigure}
    \hfill
    \begin{subfigure}[t]{0.48\linewidth}
    \centering
    \includegraphics[width=0.95\linewidth]{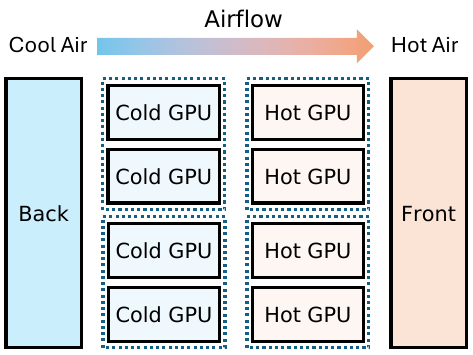}
    \caption{Airflow and cooling layout of MI250 node. GPUs within dashed box are in same package.}
    \label{fig:mi250-cooling}
    \end{subfigure}
    \vspace{-0.8em}
    \caption{Airflow and cooling system designs of the evaluated server nodes. }
    \label{fig:hgx-cooling}
    \vspace{-3ex}
\end{figure}

As illustrated in Figure~\ref{fig:hgx-cooling}(a), the NVIDIA HGX system employs front-to-back airflow, resulting in systematic cooling disparities even at idle. GPUs positioned toward the rear (near the warm air exhaust) experience consistently higher temperatures than those near the intake. 
This imbalance becomes especially problematic under compute-heavy configurations with high pipeline parallelism, where GPUs operate closer to their thermal envelope.

\begin{figure}[]
    \centering
    \begin{subfigure}[]{\linewidth}
        \centering
        \includegraphics[width=0.95\linewidth]{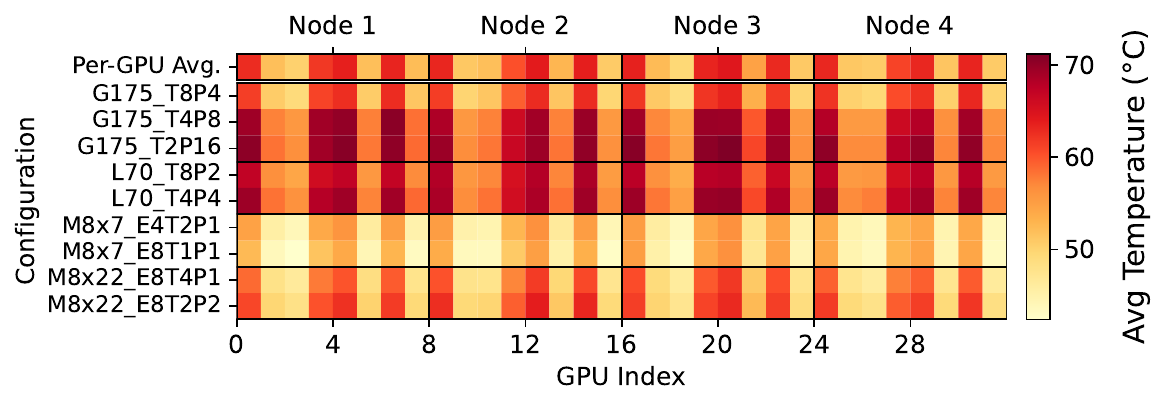}
        \vspace{-0.7em}
        \caption{Average GPU temperature heatmap.}
        \vspace{0.5em}
        \label{fig:thermal-avg}
    \end{subfigure}
    \vfill
    \begin{subfigure}[]{\linewidth}
        \centering
        \includegraphics[width=0.95\linewidth]{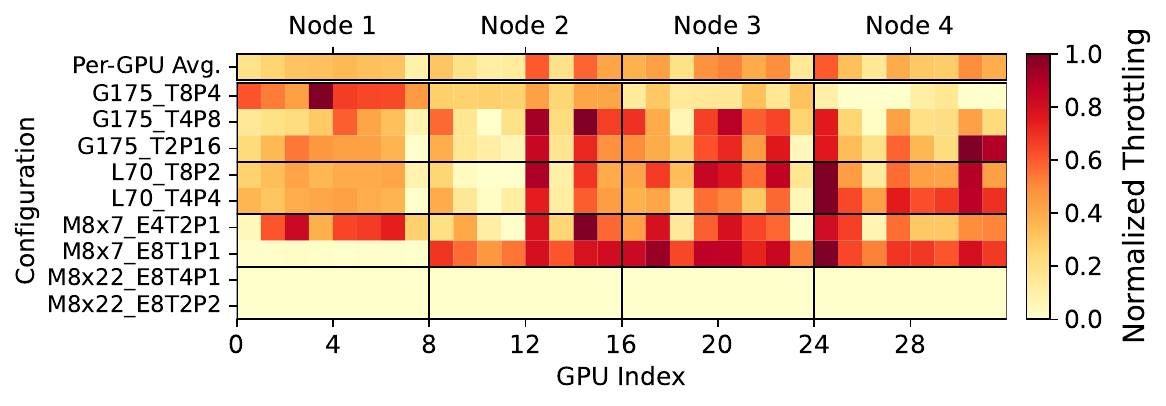}
        \vspace{-0.7em}
        \caption{Average GPU frequency throttling heatmap. Values are normalized per configuration, with the minimum value mapped to 0 and the maximum value mapped to 1 for each row. }
        \label{fig:gpu-throttling}
    \end{subfigure}
    \vspace{-1em}
    \caption{Thermal distribution and normalized clock throttling across GPUs in the H200 cluster.
    }
    \vspace{-0.5em}
    \label{fig:thermal-heatmap}
\end{figure}

\begin{figure}[]
    \centering
    \begin{subfigure}[]{\linewidth}
        \centering
        \includegraphics[width=0.95\linewidth]{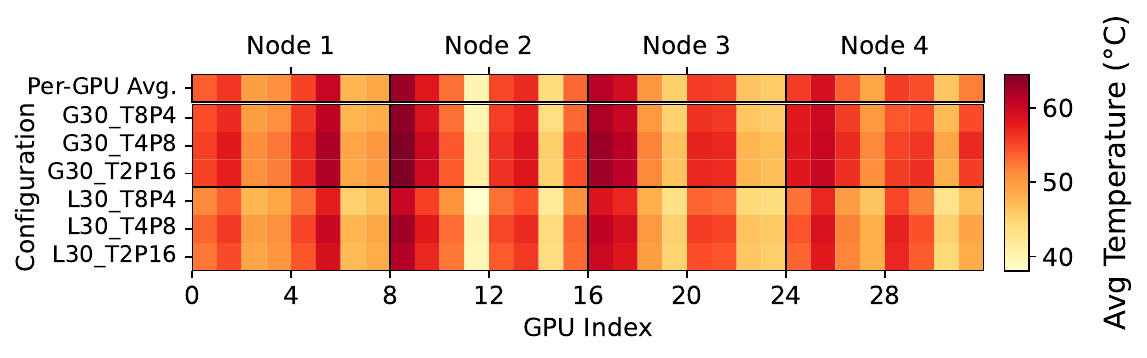}
        \vspace{-0.7em}
        \caption{Average GPU temperature heatmap.}
        \label{fig:mi250-thermal-avg}
    \end{subfigure}
    \vfill
    \begin{subfigure}[]{\linewidth}
        \centering
        \includegraphics[width=0.95\linewidth]{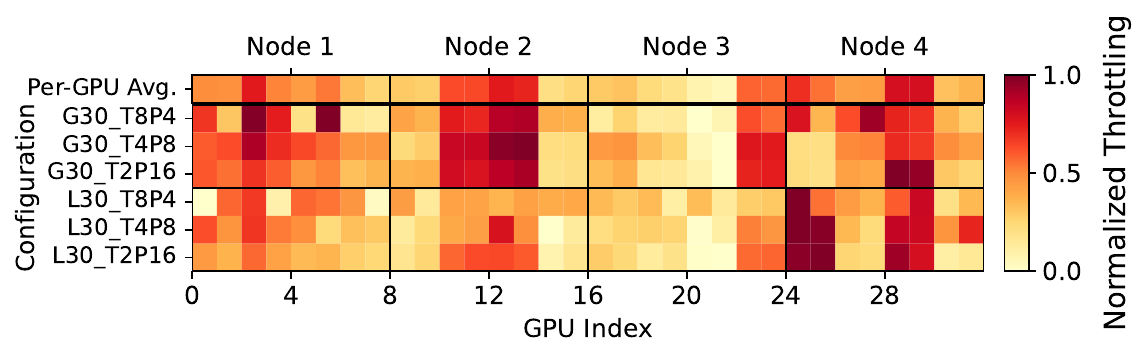}
        \vspace{-0.7em}
        \caption{Average GPU throttling heatmap. Values are normalized per configuration, ranging from 0 to 1.}
        \label{fig:mi250-gpu-throttling}
    \end{subfigure}
    \vspace{-1em}
    \caption{Thermal distribution and normalized clock throttling across GPUs in the MI250 cluster.
    }
    \vspace{-1em}
    \label{fig:mi250-thermal-heatmap}
\end{figure}

Figure~\ref{fig:thermal-heatmap} shows temperature differentials between rear and front GPUs, reaching up to 27\% in extreme cases. These elevated temperatures trigger thermal throttling in rear GPUs, as shown in Figure~\ref{fig:gpu-throttling}, where clock frequencies are reduced to prevent overheating -- ultimately lowering throughput and reducing efficiency.
A similar trend holds in AMD’s MI250 cluster. Figure~\ref{fig:mi250-thermal-heatmap} shows temperature variations even within GPU packages, due to airflow patterns and package placement, with 5–10°C temperature skew observed across paired logical GPUs. The imbalance worsens under deeper pipeline parallelism, where power consumption becomes more bursty and uneven, further stressing the cooling infrastructure.

\begin{figure}[]
    \centering
    \begin{subfigure}[t]{\linewidth}
        \centering
        \includegraphics[width=0.9\linewidth]{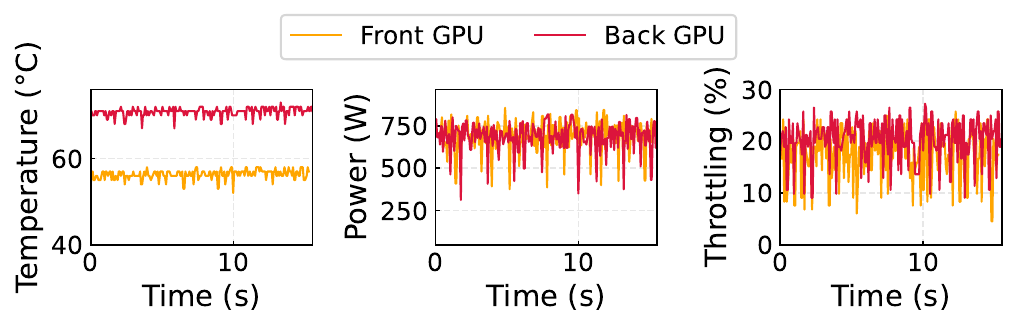}
        \vspace{-0.7em}
        \caption{Timeline of GPT3-175B training.}
        \label{fig:gpt-timeseries}
    \end{subfigure}
    \vfill
    \begin{subfigure}[t]{\linewidth}
        \centering
        \includegraphics[width=0.9\linewidth]{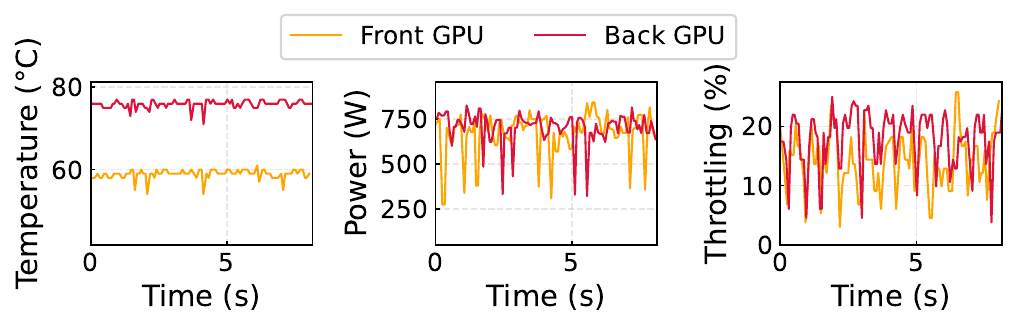}
        \vspace{-0.7em}
        \caption{Timeline of Mixtral-8x22B training.}
        \label{fig:mixtral-timeseries}
    \end{subfigure}
    \vspace{-1em}
    \caption{Thermal and performance change over time across different training workloads. Both GPT and Mixtral traces reveal consistent thermal imbalance between front and rear GPUs, leading to more frequent throttling in hotter units.}
    \label{fig:timeseries}
\end{figure}

To further analyze this, we examine the time series of power consumption and GPU temperatures during training.
Figure~\ref{fig:timeseries} reveals that power draw fluctuates across GPUs over time, with rear-positioned GPUs consistently exhibiting higher temperatures than those closer to the intake.
This imbalance remains persistent throughout the training session, with no clear cooldown periods. Sustained exposure to elevated temperatures can accelerate hardware degradation, reducing lifespan and increasing the likelihood of failures for hotter GPUs.
In addition, thermally constrained GPUs are more susceptible to clock throttling, causing them to act as stragglers in distributed execution.

To examine the impact of resource contention on thermal throttling, we co-analyze the average throttling ratio with average GPU occupancy as well as the warps and threadblocks counts.
Note that occupancy reflects the number of active warps normalized by the GPU's scheduling limits and includes both compute and communication kernels.
In contrast, the number of warps and threadblocks reflects the volume of work on the GPU, providing a more direct indication of execution pressure and potential contention.
As shown in Figure~\ref{fig:occupancy}, high-PP configurations increase threadblock and warp counts due to asynchronous execution, elevating thermal load. 
In contrast, TP setups maintain high occupancy, driven by prolonged communication kernels, but show fewer threadblocks and warps, leading to lower throttling.
Training-time optimizations further shift these metrics: activation recomputation in GPT3-175B reduces both workload pressure and occupancy, more notably in PP-heavy configurations due to elongated pipeline bubbles.
In contrast, in Llama3-70B, recomputation increases workload pressure because all configurations include DP, requiring each GPU to recompute its full share of activations independently.
Compute-communication overlap increases all three metrics as well as throttling, highlighting the trade-off between concurrency and thermal stress.
These trends reveal that thermal throttling is not caused by any single factor but rather emerges from complex interactions between physical constraints, software-level execution patterns, and hardware scheduling behaviors.

\begin{figure}
\centering
        \includegraphics[width=0.9\linewidth]{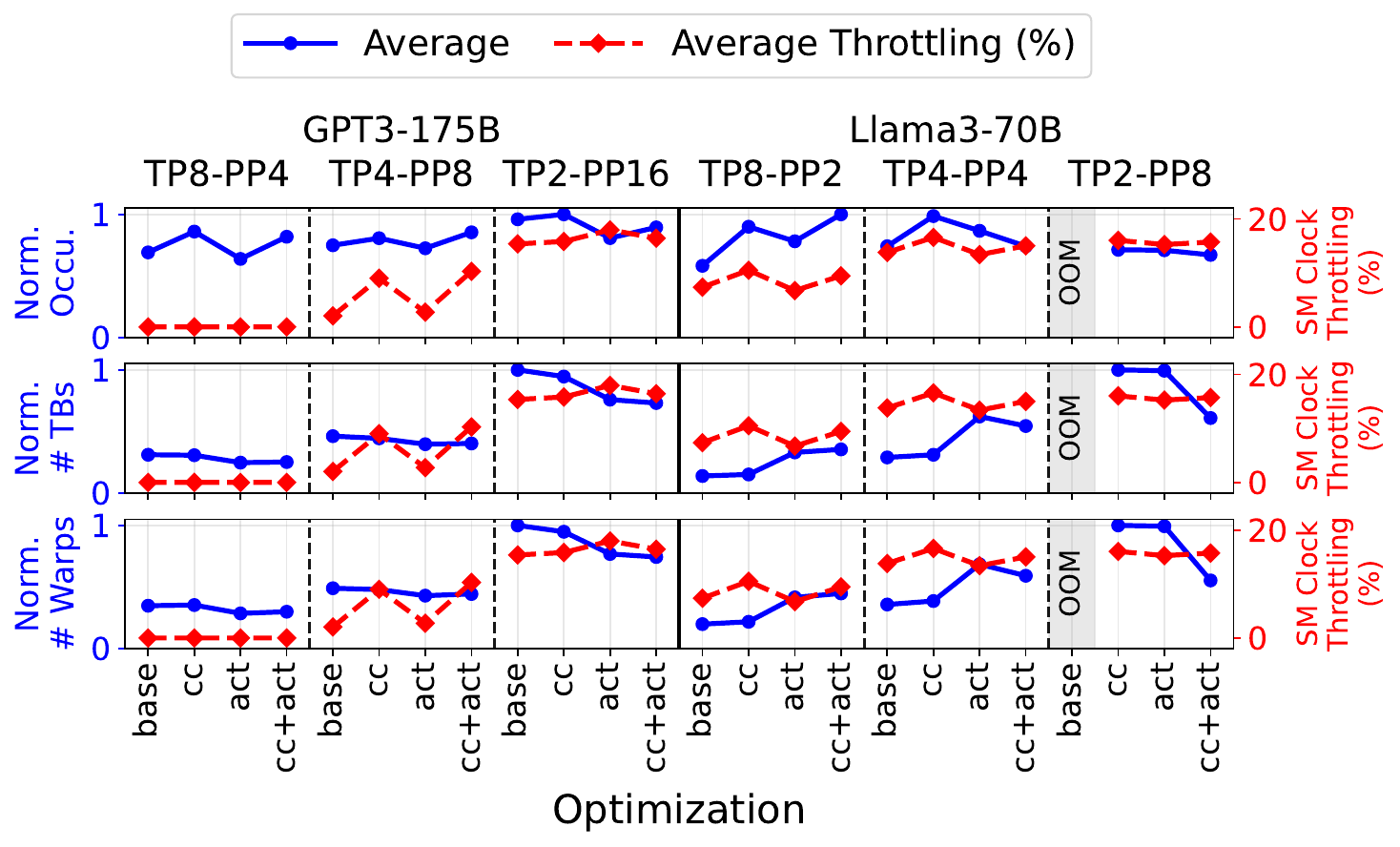}
    \label{fig:occupancy}
    \vspace{-1em}
    \caption{Average SM clock throttling and GPU hardware metrics during training on H200 cluster.}
    \vspace{-1em}
    \label{fig:occupancy}
\end{figure}

\begin{figure}
    \centering
        \includegraphics[width=0.85\linewidth]{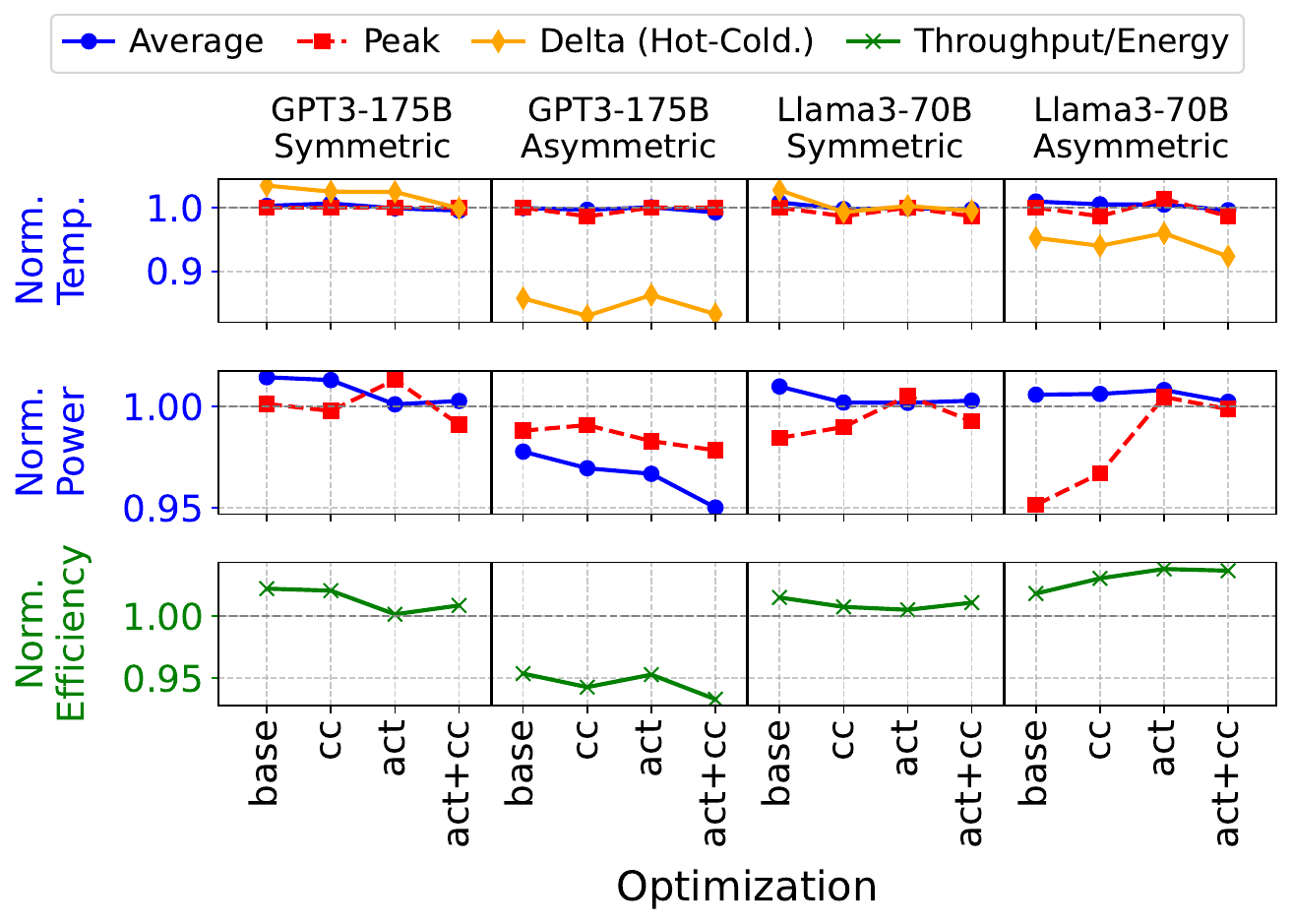}
        \label{fig:thermal-aware-pp}
    \label{fig:thermal-aware-pp}
    \vspace{-1em}
    \caption{GPU power, temperature, and training efficiency of thermal-aware pipeline stage placement, normalized by the baseline pipeline parallelism strategy. }
    \label{fig:thermal-aware-pp}
    \vspace{-3ex}
\end{figure}

A key insight from our analysis is that software systems often assume homogeneous thermal behavior and apply optimizations uniformly across GPUs.
This approach is suboptimal in thermally imbalanced systems, suggesting that cooling-aware scheduling can help mitigate thermal hotspots and improve system efficiency.
To address hardware-induced thermal imbalance, we implement a thermal-aware pipeline parallelism strategy.
Each pipeline stage consists of 4 GPUs (4-way tensor parallelism), forming 2 stages per node.
DP is disabled to ensure that each pipeline-parallel domain aligns with its corresponding thermal group.
Unlike conventional methods that group GPUs by consecutive device IDs, we cluster hot and cold GPUs into separate stages to isolate thermal effects.
Colder GPUs handle early, typically heavier stages (e.g., embedding, projection) to boost utilization and avoid throttling. 
We also apply asymmetric layer allocation, giving cooler stages an extra layer to further offload hotter GPUs.
Figure~\ref{fig:thermal-aware-pp} compares two variants: Symmetric (cold GPUs on early stages) and Asymmetric (adds layer imbalance).
The symmetric setup improves efficiency by up to 2\% via better load balancing, but increases thermal variance and power draw.
Asymmetric results vary by model granularity.
Llama3-70B (80 layers, 4 stages) allows a 19/21 split (10\% imbalance), yielding a 4\% efficiency gain and 8\% temperature gap reduction. 
In contrast, GPT3-175B (96 layers, 8 stages) has an 11/13 split (18\% imbalance), reducing thermal gap by 17\% but degrading efficiency by 7\%.
These results show that, when aligned with model and hardware characteristics, thermal-aware scheduling can improve both efficiency and thermal behavior.

\vspace{0.5em}
\noindent\fbox{%
    \parbox{\columnwidth}{%
    \textit{\textbf{Insight: Thermal imbalance across GPUs impacts system reliability and performance.} 
GPUs positioned near the exhaust end of the chassis consistently reach higher temperatures due to airflow limitations, resulting in persistent thermal imbalance. This leads to increased clock throttling, reduced training efficiency, and accelerated hardware aging. Cooling-aware strategies, such as thermal-aware workload scheduling, can mitigate some of the effects by balancing the thermal load, thereby improving both performance and long-term reliability.}}}%

\vspace{1em}

\section{Discussion}\label{sec:discussion}

\subsection{Extending to Datacenter-Scale Systems}\label{subsec:extrapolation}

To further our analysis at larger scales, we model GPT-3 175B training on clusters with up to 8K GPUs using Astra-Sim~\cite{astrasim, astrasim2} as the network simulator, combined with profiling data from real GPUs presented in the previous sections.
We project performance by increasing the degree of data parallelism while maintaining the same tensor and pipeline parallelism, resulting in a total GPU count of TP$\times$PP$\times$DP.
For DP=1, we directly use real-system kernel latencies measured on 32$\times$H200 and 64$\times$H100 clusters.
To project performance for larger DP values, we first divide compute and communication time by the DP degree, then add the modeled DP AllReduce time to estimate overall performance.
When modeling inter-node bandwidth scaling, we divide inter-node communication time by the bandwidth multiplier (e.g., $\times$8) and add the modeled AllReduce time across DP degrees.

\begin{figure}
\centering
    \begin{subfigure}[t]{\linewidth}
        \centering
        \includegraphics[width=1\linewidth]{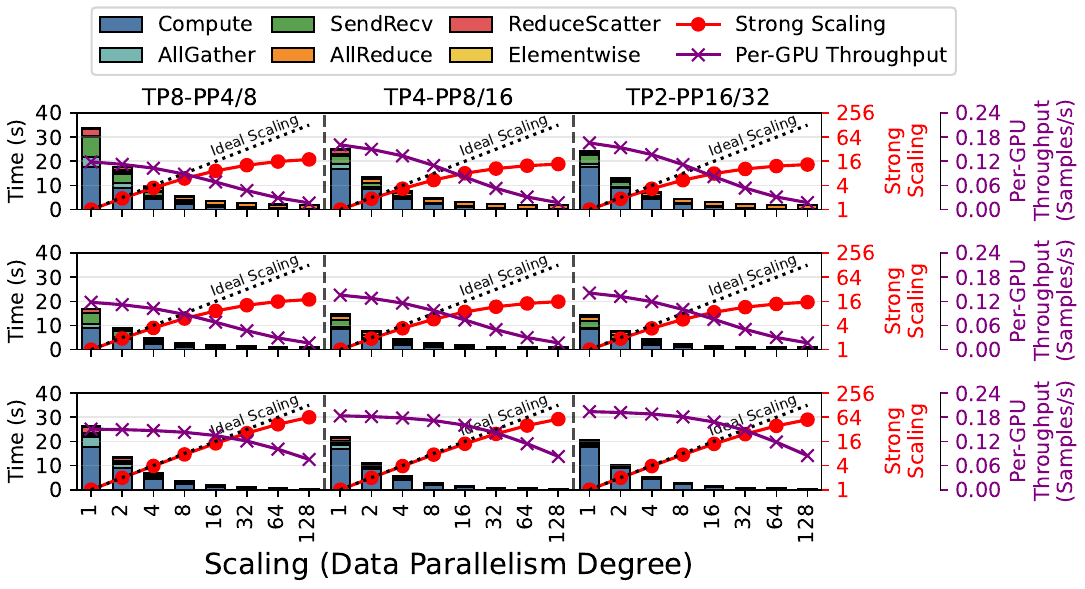}
        \label{fig:extrapolation}
    \end{subfigure}
    \vspace{-2.2em}
    \caption{Projected per-kernel latency, strong scaling, and per-GPU throughput for different model parallelism configurations and system scales on H200 cluster (top) and H100 cluster (center) with 100G inter-node bandwidth, and H200 cluster with 800G inter-node bandwidth (bottom). }
    \label{fig:extrapolation-throughput}
    \vspace{-0.5em}
\end{figure}

Figure~\ref{fig:extrapolation-throughput} breaks down projected latency by kernel, strong scaling efficiency, and per-GPU throughput across H200 and H100 clusters under different interconnect bandwidths.
We observe that H100 clusters achieve higher absolute throughput than H200 clusters, but lower per-GPU throughput due to increased communication overhead.
This highlights the efficiency advantage of H200 clusters, which benefit from larger per-device memory capacity that reduces communication overhead.
Our results further demonstrate that naive scaling yields sublinear throughput improvements due to communication bottlenecks.
At 100Gbps bandwidth, both H200 and H100 clusters suffer from significant AllReduce overhead at larger DP degrees,  with strong scaling dropping by up to 9.7$\times$ compared to the ideal case, revealing persistent scaling limits.
This demonstrates that while larger DP degrees enable higher batch sizes, they also increase communication bandwidth requirements as the system scales, making network performance an even more critical factor.
To assess the impact of network bandwidth, we also project performance assuming 800Gbps interconnects.
In this setting, the strong scaling efficiency improves as much as 4.2$\times$ compared to the 100Gbps baseline.
This shows that as the system scales, higher-bandwidth interconnects can significantly alleviate communication bottlenecks, enabling much closer-to-ideal scaling for large DP configurations.
These findings strengthen our insights -- the need for co-optimizing parallelization strategies, hardware capabilities, and system topology rather than assuming uniform gains from each dimension in isolation.

\begin{figure}[t]
    \centering
    \includegraphics[width=0.95\linewidth]{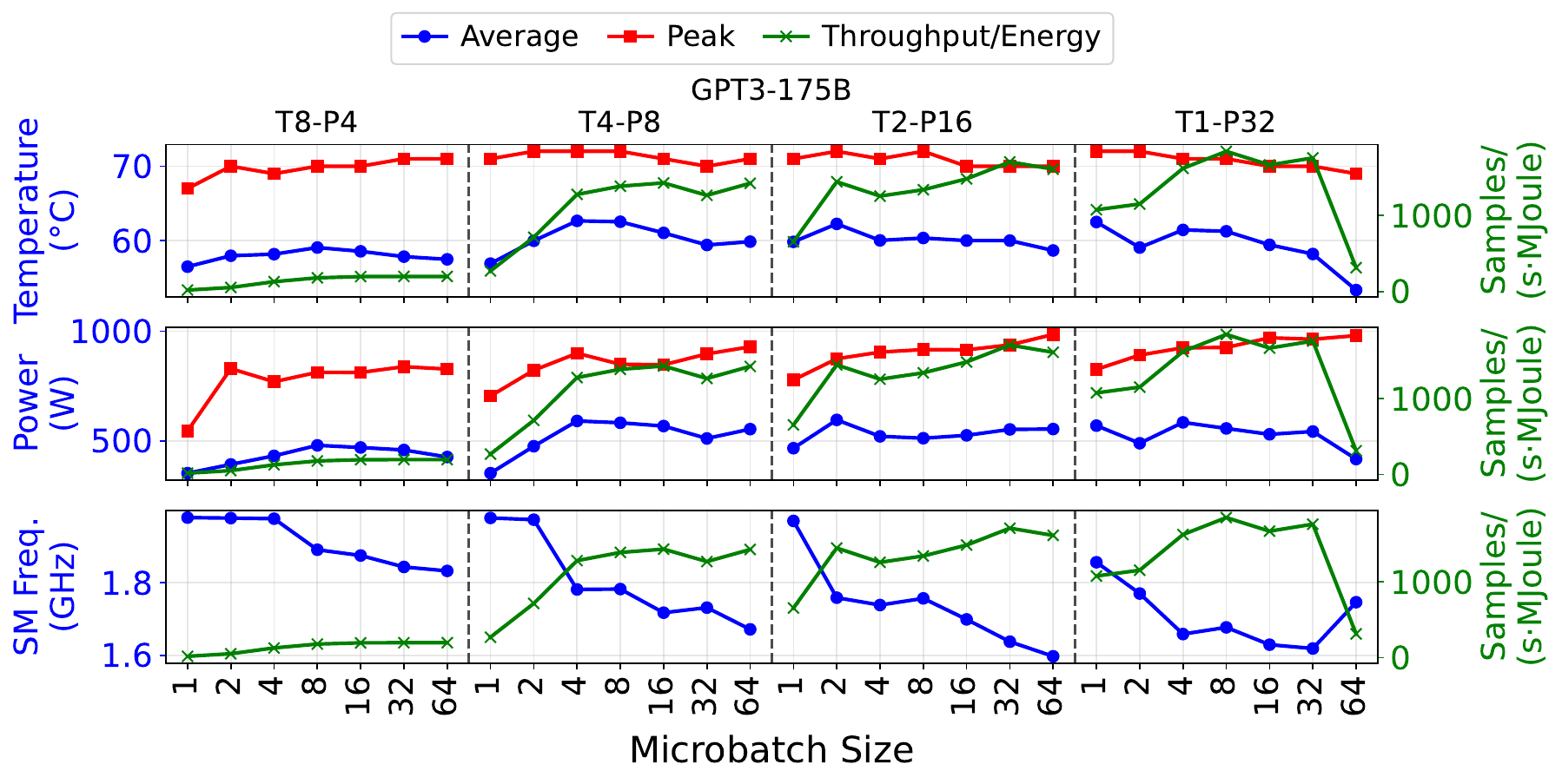}
    \vspace{-1em}
    \caption{GPU power, temperature, and clock frequency during inference on the H200 cluster, across parallelism configs, and microbatch sizes.}
    \label{fig:ubatch-sweep-inference}
\end{figure}

\vspace{0.3em}

\subsection{Characterizing Distributed Inference}\label{subsec:inference}

Unlike training, where communication bottlenecks often dominate, inference involves less inter-GPU communication since model weights are fixed.
To study how microbatch size affects distributed inference, we sweep a range of microbatch sizes across various parallelism settings.
Figure~\ref{fig:ubatch-sweep-inference} shows that larger microbatches generally improve throughput without significantly increasing average power or temperature. This is likely due to fewer synchronization steps and lower communication volume.
We also observe that inference consumes less average power and produces lower thermal loads than training, reflecting its lower computational intensity. 
However, peak power and temperature remain high due to bursty operations like attention and large matrix multiplications.
These dynamics highlight the complexity of real-world inference, which must accommodate dynamic request rates, model switching, and variable latency demands.
In such settings, thermal-aware schedulers can potentially improve performance by routing latency-sensitive or compute-intensive tasks to cooler GPUs, or by strategically assigning heavier stages in distributed or disaggregated inference.

\vspace{0.3em}

\subsection{Maintaining Efficiency by Optimizing for Physical Realities of ML System}\label{subsec:futurework}

Our findings highlight a growing need to broaden the scope of optimization beyond isolated algorithmic and parallelism strategies to encompass hardware-level metrics and constraints.
Hardware scaling, interconnect asymmetry, and thermal imbalance are no longer secondary effects due to the scale of modern large-scale ML training.
Even in modern datacenter infrastructure with advanced cooling designs~\cite{meta2024sustainability,dgxcooling}, temperature skew persists due to residual airflow mixing, rack constraints, and workload-
induced heat concentration~\cite{stojkovic2024dynamollm, tapas, googleBlog}.
As our analyses show, these factors can meaningfully impact the training efficiency, performance and system reliability, even in a well-balanced distributed workload such as LLMs.
Conventional strategies that assume uniform hardware behavior often lead to only-partially true belief that runtime performance is also uniform. As a result, often prior works advocate one-size-fits-all optimizations—such as always overlapping compute and communication, uniformly increasing microbatch sizes, or applying aggressive pipelining. However, our analysis shows that such assumptions break down in practice. Variations in power availability, thermal profiles, and clock throttling create substantial runtime heterogeneity across devices, even for the same hardware.

To address this, we argue for systems-level co-optimization that accounts for these physical realities. This includes cooling-aware workload placement that considers airflow patterns, thermal- and power-aware scheduling policies that adapt dynamically to temperature and utilization, adaptive microbatch scaling to match device performance, and system infrastructure capable of detecting and responding to power, frequency, and performance anomalies in real time.
These interventions are essential to sustaining training efficiency and system robustness as models and clusters scale.

\vspace{-0.5em}
\section{Conclusion}

This work highlights key system-level bottlenecks in large-scale LLM training that go beyond compute and memory. We show that increasing microbatch size doesn’t always lead to better performance due to communication bursts and thermal stress. Thermal imbalance across GPUs, driven by airflow and power delivery limits, causes clock throttling and hardware strain, reducing training efficiency.
Our characterization and analysis call for more hardware- and thermals-aware scheduling, rather than relying on uniform optimization assumptions. Future systems must co-optimize placement, parallelism, power, and thermal to maximize both performance and reliability at scale.
\begin{acks}

This research was supported in part through research cyberinfrastructure resources and services provided by the AI Makerspace and the Partnership for an Advanced Computing Environment (PACE) at the Georgia Institute of Technology, Atlanta, Georgia, USA, and by the AMD HPC Fund Research Cloud. 
We thank Eric Coulter and Michael Weiner for their assistance in provisioning access to the Georgia Tech AI Makerspace~\cite{aimakerspace,forbes}. 
This work was partially supported by gifts from Google and AMD, and by ACE, one of seven centers in JUMP 2.0, a Semiconductor Research Corporation (SRC) program sponsored by DARPA. 
We thank Matthieu Bloch for enabling this research. 
The views and conclusions contained herein are those of the authors and should not be interpreted as representing the official policies or endorsements, either expressed or implied, of Georgia Tech.

\end{acks}
%
%
%
%
%






\appendix
\section{Artifact Appendix}

\subsection{Abstract}

The artifact includes the source code of the evaluated frameworks and a modified Zeus implementation for hardware telemetry collection. It also includes SLURM and bash scripts to prepare and launch experiments, along with Python scripts to reproduce and visualize key results from the paper. The artifact is publicly available on Github (\textcolor{ACMDarkBlue}{\url{https://github.com/sitar-lab/CharLLM-PPT}}) and Zenodo (\textcolor{ACMDarkBlue}{\url{https://doi.org/10.5281/zenodo.16734738}}).

\subsection{Artifact check-list (meta-information)}

{\small
\begin{itemize}
  \item {\bf Program:} NVIDIA NeMo, Megatron-LM and dependencies
  \item {\bf Compilation:} gcc, nvcc, cmake, ninja
  \item {\bf Model:} GPT3 175B, Llama3 70B, Mixtral 8x7B, Mixtral 8x22B
  \item {\bf Data set:} The Pile
  \item {\bf Run-time environment:} RHEL 9 with SLURM and Anaconda 3
  \item {\bf Hardware:} Training cluster with 32 NVIDIA H200 GPUs
  \item {\bf Metrics:} Training throughput and efficiency, power consumption, temperature, and network traffic
  \item {\bf Output:} System telemetry CSV files, Chakra traces, and visualization figures
  \item {\bf How much disk space required (approximately)?:} 1TB
  \item {\bf How much time is needed to prepare workflow (approximately)?:} \textasciitilde10 hours
  \item {\bf How much time is needed to complete experiments (approximately)?:}  5\textasciitilde6 days (if executed serially)
  \item {\bf Publicly available?: }Yes
  \item {\bf Code licenses (if publicly available)?:} MIT License
  \item {\bf Archived (provide DOI)?:} \href{https://doi.org/10.5281/zenodo.16734738}{[\textcolor{ACMDarkBlue}{https://doi.org/10.5281/zenodo.16734738}]}
\end{itemize}
}

\subsection{Description}

\subsubsection{How to access}
All source code and scripts are packaged into a single tarball archive. The artifact can be downloaded from the archive link provided.

\subsubsection{Hardware dependencies}
Experiments were conducted on four HGX H200 nodes connected via InfiniBand switches with 100 Gbps inter-node bandwidth.

\subsubsection{Software dependencies}
RHEL 9 or similar, SLURM workload manager, and Anaconda 3. NVML and Prometheus Node Exporter are required for hardware telemetry collection.

\subsubsection{Data Sets}
We use the Pile training dataset, publicly available via 
\href{https://huggingface.co/datasets/monology/pile-uncopyrighted}{\textcolor{ACMDarkBlue}{HuggingFace}}.

\subsection{Installation and Testing}

\subsubsection{Installation}
{\small
\begin{verbatim}
# Create conda environment (name must match)
conda create -n AIInfraAnalysis-AE python=3.10 -y
conda activate AIInfraAnalysis-AE

# Set project root directory
export AIINFRA_ROOT=<PATH_TO_PROJECT_ROOT>

# Install prerequisites (may take ~10 hours)
cd $AIINFRA_ROOT


# (Download the artifact from the DOI to $AIINFRA_ROOT)
tar -xvf aiinfra-artifact.tar.gz
cd AIInfraAnalysis/scripts/
./install.sh
\end{verbatim}
}

\subsubsection{Basic Test}
{\small
\begin{verbatim}
# Set project root directory
export AIINFRA_ROOT=<PATH_TO_PROJECT_ROOT>

# Modify prepare_test.sh to match your environment then run it
vi ${AIINFRA_ROOT}/AIInfraAnalysis/scripts/prepare_test.sh
bash ${AIINFRA_ROOT}/AIInfraAnalysis/scripts/prepare_test.sh

# Launch the test experiment (may take ~1 hour)
bash ${AIINFRA_ROOT}/AIInfraAnalysis/scripts/launch_test.sh

# Logs, CSV files, and Chakra traces will be generated under: 
# ${AIINFRA_ROOT}/AIInfraAnalysis/results/test
\end{verbatim}
}

\subsection{Experiment Workflow}
The preparation script generates SBATCH job scripts based on user-specified configuration. Once the scripts are ready, the launcher script submits all experiments to the SLURM workload manager.

\begin{enumerate}
  \item Modify \texttt{AIInfraAnalysis/scripts/prepare\_scripts.sh} to match your SLURM environment, then execute it.
  \item Run \texttt{AIInfraAnalysis/scripts/prepare\_datasets.sh} to download all required datasets.
  \item Run \texttt{AIInfraAnalysis/scripts/full\_sweep.sh} to launch all experiments.
  \item To launch a subset of the experiments, refer to the instructions in the \texttt{README.md} file.
  \item After completion, generate the following paper figures using scripts in 
  \texttt{AIInfraAnalysis/scripts/visualization/}:
  \begin{itemize}
    \item Figure 4: \texttt{figure\_4.sh}
    \item Figures 5 and 17: \texttt{figure\_heatmaps.sh}
    \item Figure 8: \texttt{figure\_8.sh}
    \item Figure 10: \texttt{figure\_10.sh}
    \item Figure 13: \texttt{figure\_13.sh}
    \item Figure 15: \texttt{figure\_15.sh}
  \end{itemize}
\end{enumerate}

\subsection{Evaluation and Expected Results}
This artifact reproduces the key results presented in Figures 4, 5, 8, 10, 13, 15, and 17 of the paper. All data generated during evaluation will be stored in \texttt{AIInfraAnalysis/results/}, and the corresponding figures will be saved in \texttt{AIInfraAnalysis/figures/}. 
Due to the nature of real-system experiments, slight variations in training time and hardware telemetry measurements are expected.




\balance
\bibliographystyle{ACM-Reference-Format}
\bibliography{reference}

\end{document}